\documentclass[english,american]{IEEEtran}
\usepackage[LGR,T1]{fontenc}
\usepackage[utf8]{inputenc}
\usepackage{color}
\usepackage{babel}
\usepackage{array}
\usepackage{float}
\usepackage{units}
\usepackage{multirow}
\usepackage{amstext}
\usepackage{amsthm}
\usepackage{graphicx}
\usepackage[unicode=true]
 {hyperref}

\makeatletter

\newcommand{\noun}[1]{\textsc{#1}}
\DeclareRobustCommand{\greektext}{%
  \fontencoding{LGR}\selectfont\def\encodingdefault{LGR}}
\DeclareRobustCommand{\textgreek}[1]{\leavevmode{\greektext #1}}

\providecommand{\tabularnewline}{\\}
\floatstyle{ruled}
\newfloat{algorithm}{tbp}{loa}
\providecommand{\algorithmname}{Algorithm}
\floatname{algorithm}{\protect\algorithmname}

\theoremstyle{plain}
\newtheorem{thm}{\protect\theoremname}
\theoremstyle{remark}
\newtheorem{rem}[thm]{\protect\remarkname}
\theoremstyle{definition}
\newtheorem{problem}[thm]{\protect\problemname}
\theoremstyle{definition}
\newtheorem{sol}[thm]{\protect\solutionname}

\@ifundefined{date}{}{\date{}}
\usepackage{algorithmic}
\usepackage{cite}

\ifdefined\showcaptionsetup
 \PassOptionsToPackage{caption=false}{subfig}
\fi
\usepackage{subfig}
\makeatother

\addto\captionsamerican{\renewcommand{\problemname}{Problem}}
\addto\captionsamerican{\renewcommand{\remarkname}{Remark}}
\addto\captionsamerican{\renewcommand{\solutionname}{Solution}}
\addto\captionsamerican{\renewcommand{\theoremname}{Theorem}}
\addto\captionsenglish{\renewcommand{\algorithmname}{Algorithm}}
\addto\captionsenglish{\renewcommand{\problemname}{Problem}}
\addto\captionsenglish{\renewcommand{\remarkname}{Remark}}
\addto\captionsenglish{\renewcommand{\solutionname}{Solution}}
\addto\captionsenglish{\renewcommand{\theoremname}{Theorem}}
\providecommand{\problemname}{Problem}
\providecommand{\remarkname}{Remark}
\providecommand{\solutionname}{Solution}
\providecommand{\theoremname}{Theorem}

\begin{document}
\title{A Tutorial on Controlling Metasurfaces from the\\
Network Perspective}
\author{Christos Liaskos$^{1}$, Evangelos Papapetrou$^{2}$, Kostas Katsalis$^{3}$,
Dimitrios Tyrovolas$^{4}$, Alexandros Papadopoulos$^{5}$, Stavros
Tsimpoukis$^{6}$, Arash~Pourdamghani$^{7}$, Max~Franke$^{8}$,
Stefan Schmid$^{9}$\\
{\small$^{1,2,5,6}$University of Ioannina, Ioannina, Greece, emails:
\{cliaskos, epap, a.papadopoulos, s.tsimpoukis\}@uoi.gr}\\
{\small$^{3}$DOCOMO Euro-Labs, Munich, Germany, email: Katsalis@docomolab-euro.com}\\
{\small$^{4}$University of Patras, Greece, email: dtyrovolas@upnet.gr}\\
{\small$^{7,8,9}$Technical University of Berlin, Germany, emails:
\{pourdamghani, m.franke, stefan.schmid\}@tu-berlin.de.}}
\maketitle
\begin{abstract}
Metasurfaces have emerged as transformative electromagnetic structures
for wireless communications, enabling the real-time control over wave
propagation, yielding potential for improved data rates, privacy,
energy efficiency and even precise environmental sensing. This tutorial
offers a perspective on controlling metasurfaces by treating them
as components of a larger networked system. Towards this end, we first
review the physical principles of metasurfaces and their various applications,
followed by an exploration of manufacturing approaches for creating
these structures. Then, aligning with standard network-layer concepts,
we describe the modeling of metasurfaces as wave routers, enabling
us to describe systems of metasurfaces using graph theory. This approach
enables the development of a performance objective framework for optimizing
these systems, while classes of heuristic and path-finding-driven
algorithms are discussed as practical solvers. The paper also examines
the integration of metasurfaces with communication systems, by presenting
their overall workflow, discussing its relation to ongoing standardization
efforts, as well as defining a context for their integration to network
simulators, using Omnet++ as a driving example. Finally, the paper
explores future directions for research in this field, identifying
graph-theoretic, standardization and integration challenges, relating
to several networking disciplines, such as time-sensitive networking,
age of information and scheduling. The paper also considers the potential
of AI-driven applications, beyond classic data transfer objectives.
\end{abstract}


\section{Introduction}\label{sec:Introduction}

In recent years, wireless networks have begun to undergo a paradigm
shift, extending connectivity beyond user devices to include previously
passive obstacles in the environment~\cite{ref1}. The
enablers of this paradigm are known as metasurfaces, thin artificial
materials supporting programmable interaction types with impinging
wireless waves~\cite{ref2}. The interactions include programmable
steering, focusing, absorption, phase control, and polarization manipulation. 

Leveraging such functionalities in a system where planar objects have
been covered with metasurfaces, denoted as Programmable Wireless Environments
(PWEs)~\cite{ref3}, has allowed for considerable performance
gains in network data rates, privacy, coverage and energy efficiency.
These gains result from configuring a PWE to mitigate the long-standing
degradation factors of wireless networks. Signal dissipation can be
countered by wave focusing, blockage can be countered by steering
waves via alternative paths, a technique which can also be exploited
for improving privacy and reducing interference. 

Historically, PWEs constitute the culmination of an ongoing effort
to exert control over the propagation process. Preceding efforts included
the use of passive relays~\cite{ref4,ref5}, large
antenna arrays~\cite{ref6}, and metasurfaces~\cite{ref7},
evolving from coarse-grained to fine control over waves, respectivelly,
within a strict signal processing context. In the PWE iteration, the
control over wireless propagation was structured over networking principles~\cite{ref8}.
One significant advantage of this approach is the ability to optimize
metasurface systems by considering them as part of a broader network
of interconnected systems and services, while adopting an OSI-compliant
approach for their operation and future evolution. 

However, while the introduction of metasurfaces to the discipline
of wireless communications has produced significant outcomes, the
multi-disciplinary nature of the field and the abundance of research
output calls for a structured tutorial aimed at an audience with a
networking background. To this end, this paper presents a perspective
on controlling metasurfaces by treating them as nodes of a network.
By leveraging concepts from network practices and graph theory, we
demonstrate how this approach can provide new insights into the design,
optimization, and integration of metasurfaces with existing infrastructure.
Specifically, the contributions of the present paper are as follows:
\begin{itemize}
\item We provide the necessary prerequisites from the aspect of physics,
electronics, resource utilization and the related nomenclature in
a compact and self-complete manner, targeting computer and network
scientists (Section~\ref{sec:Background:-From-Metasurface}).
\item Employing network-layer concepts, we detail a model for representing
metasurfaces as wave routers and systems metasurfaces, i.e., PWEs,
as graphs, and their physical capabilities as network functions (Section~\ref{sec:Metasurfaces-as-Network}).
\item We formulate the problem of configuring PWEs for a set of users as
a resource sharing optimization problem, relating to the existing
concepts of network orchestration and network update consistency (Section~\ref{sec:PWE-Performance-Optimization}). 
\item We discuss the integration of PWEs to existing networks by first defining
their workflow from initialization to operation, and then exploring
the existing standardization landscape and its future directions (Section~\ref{sec:The-PWE-integration}). 
\item We provide a context for simulating PWEs using well-known network
simulators, using Omnet++ as the driving example (Section~\ref{sec:Simulating-Networked-PWEs:}). 
\item We present an exemplary, simulation-driven evaluation scenario employing
PWEs in a challenging wireless optimization, involving the mitigation
of Doppler spread in a factory setting (Section~\ref{sec:Evaluation:-A-Factory}). 
\item We relate future challenges to existing networking directions such
as time-sensitive networking, age of information, scheduling, traffic
engineering and routing (Section~\ref{sec:Discussion-and-Network}). 
\end{itemize}
Notably, the merge of metasurfaces and communications has been highly
prolific in terms of research outcomes in recent years. This paper
is not intended as an extensive survey of these numerous efforts,
but rather seeks to serve as a tutorial and a roadmap for network-layer
and computer theory researchers interested in exploring the intersection
of metasurfaces and their respective field. 

\section{Background: From Metasurface Physics to Modeling as Network Components}\label{sec:Background:-From-Metasurface}

\begin{table*}
\caption{Terminology\textcolor{magenta}{{} }}\label{tab:Terminology}

\begin{tabular*}{1\textwidth}{@{\extracolsep{\fill}}|>{\centering}p{0.25\textwidth}|>{\centering}p{0.7\textwidth}|}
\hline 
\textbf{Term} & \textbf{Definition}\tabularnewline
\hline 
\hline 
Software-Defined Metamaterial (SDM) & General term for all planar and non-planar metamaterials that can
manipulate impinging electromagnetic waves following software commands,
to control wireless signal propagation.\tabularnewline
\hline 
Metasurface & The most advanced design approach for planar SDMs at the physical
layer (i.e., not encompassing software or control circuitry). Metasurfaces
enable the manipulation of electromagnetic waves in almost any desired
way.\tabularnewline
\hline 
Reconfigurable Intelligent Surface (RIS) & A specific SDM technology, that employs reflectarrays at the physical
layer, offering a subset of the metasurface capabilities.\tabularnewline
\hline 
Tile & Any planar SDM, when deployed in batches to cover a large surface,
such as a wall.\tabularnewline
\hline 
Programmable Wireless Environment (PWE) & A 3D space whose inner surfaces are covered with SDMs, either in part
or in full, allowing for the propagation of wireless waves to be programmable
by assigning EM Functions to tiles.\tabularnewline
\hline 
 & A template describing an EM manipulation type and its parameters.
(E.g., STEER a wave, by defining incoming and reflection directions).\tabularnewline
\hline 
Metasurface codebook & A database containing the circuit-level metasurface-internal instructions
to obtain every supported .\tabularnewline
\hline 
\end{tabular*}
\end{table*}
In this section we provide a short survey of how metamaterials look,
are manufactured and operate from the standpoint of physics. The goal
is to provide the basic foundations for understanding their network-layer
modeling, which constitutes the main goal of this paper, and is provided
in Section~\ref{sec:Metasurfaces-as-Network} and on.

\subsection{Metasurfaces: A Review of Their Physical Principles and Applications}

Metasurfaces, also known as 2D meta-materials or artificial surfaces,
have revolutionized various fields of research and technology due
to their unique electromagnetic properties and potential applications~\cite{ref9}.
From a physical perspective, metasurfaces are engineered structures
composed of periodic arrays of sub-wavelength elements that interact
with incident electromagnetic waves. This section surveys the underlying
physics behind metasurfaces and their practical applications, to provide
a basic level of understanding required for detailing the network
control aspect.

\begin{figure}[!t]
\begin{centering}
\subfloat[\label{fig:Photographic-sample-of}Photographic sample of a $\nicefrac{\lambda}{8}$
metasurface (front-side), manufactured as a printed circuit board.
(EU Project VISORSURF, \protect\href{https://projects.research-and-innovation.ec.europa.eu/en/projects/success-stories/all/hypersurfaces-control-electromagnetic-energy-app}{https://projects.research-and-innovation.ec.europa.eu/en/projects/success-stories/all/hypersurfaces-control-electromagnetic-energy-app}).]{\begin{centering}
\includegraphics[width=0.9\columnwidth]{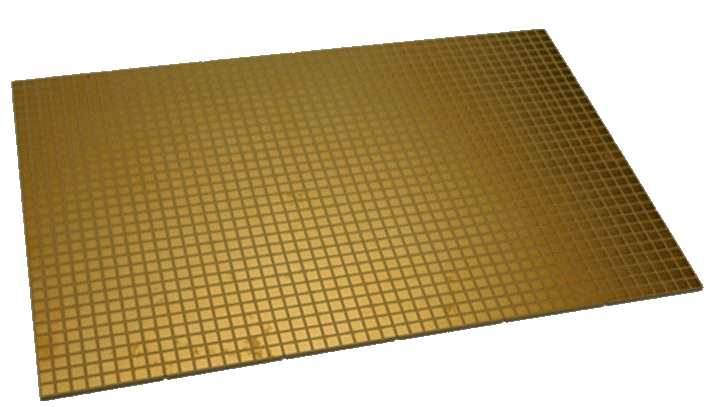}
\par\end{centering}
}
\par\end{centering}
\begin{centering}
\subfloat[\label{fig:Front-side-schematic-close}Front-side schematic close
up and backside components.]{\begin{centering}
\includegraphics[width=1\columnwidth]{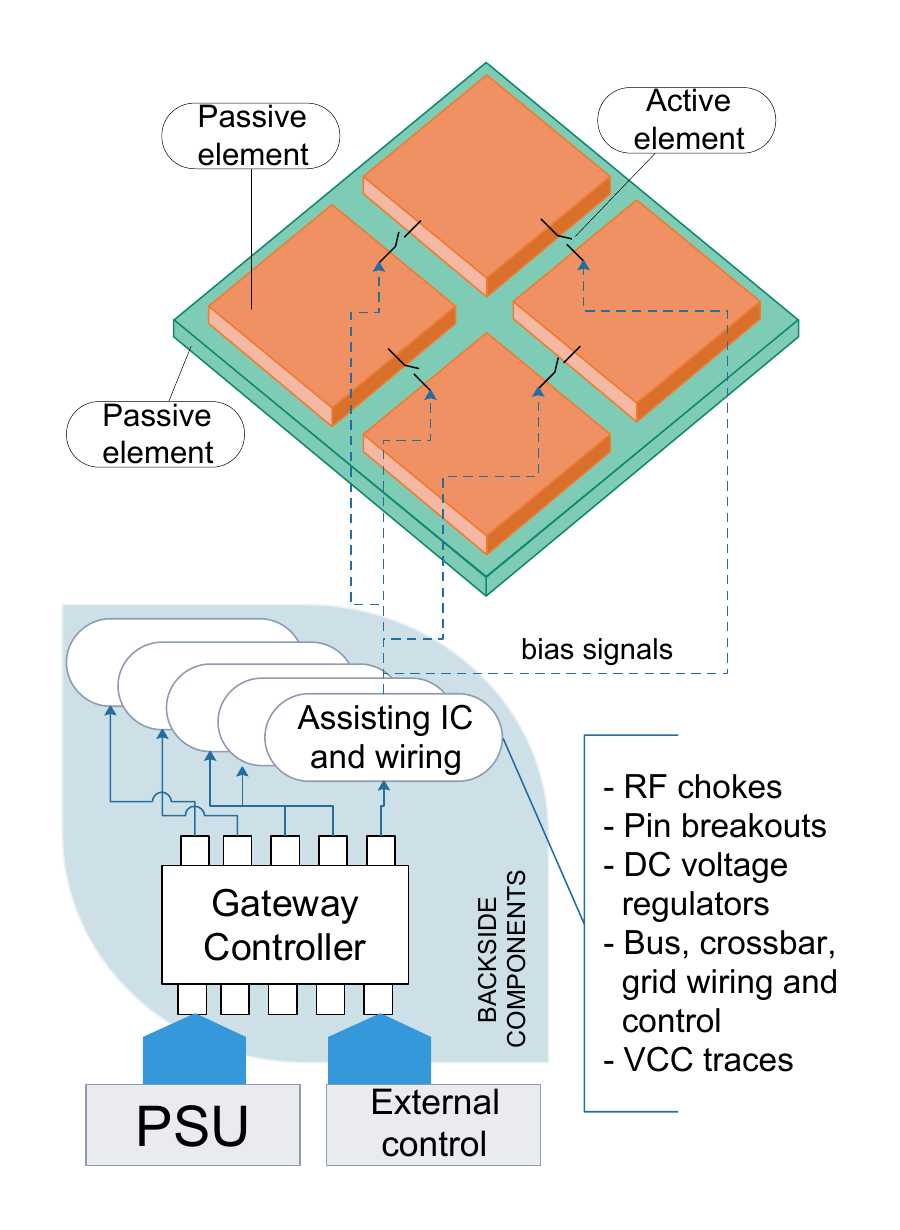}
\par\end{centering}
} 
\par\end{centering}
\caption{Typical constituents of a metasurface.}\label{fig:MSconstituents}
\end{figure}

\paragraph*{Typical appearance}

Metasurfaces commonly (but not exclusively) consist of an array of
a metal-on-dielectric basic design arranged in a periodic pattern~\cite{ref10},
as shown in Fig.~\ref{fig:MSconstituents}. This basic design is
usually denoted as \emph{cell} and can also comprise components with
controllable impedance in the radio-frequency (RF) regime. In this
case, the metasurface is deemed as \emph{programmable} and a separate,
DC-logic circuitry transfers control over the tunable impedances to
an external computer. Special circuit components, known as \emph{RF
chokes}, ensure that no RF signal leaks into the DC-logic circuitry.

The dimensions of a cell is at least in the $\nicefrac{\lambda}{2}$
order and, ideally, go below $\nicefrac{\lambda}{8}-\nicefrac{\lambda}{10}$,
$\lambda$ being the wavelength of the incident electromagnetic wave,
which yields a significant degree of modification potential of the
reflective behavior of the metasurface~\cite{ref11}. This includes achieving
negative refractive index, perfect absorption, and customized scattering.
These properties arise from the interaction between the incident wave,
the metasurface's periodic elements, and the chosen impedance values,
which can either reinforce or cancel out specific frequency components
of the wave.

\begin{figure*}[!t]
\begin{centering}
\includegraphics[width=1\textwidth]{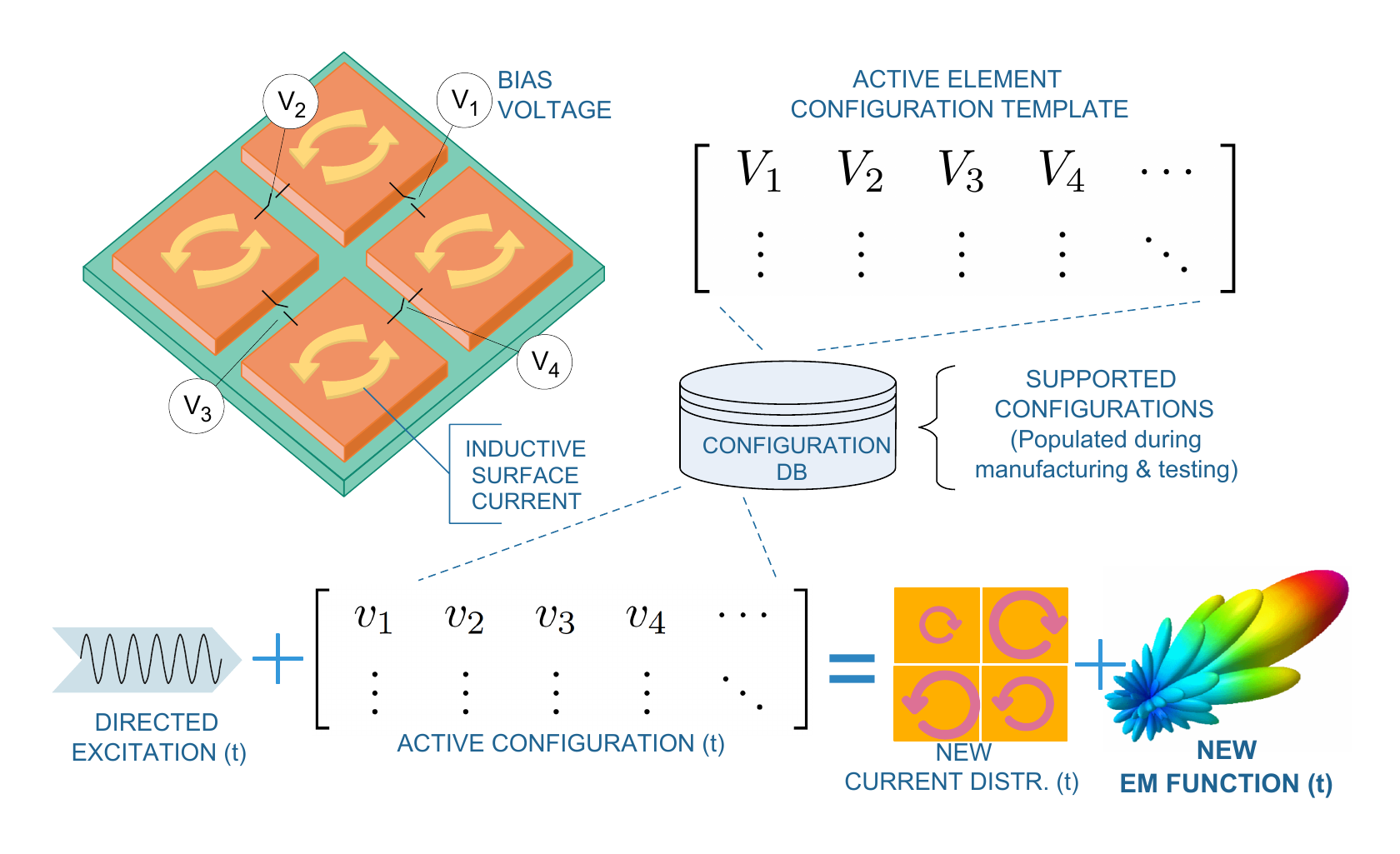} 
\par\end{centering}
\caption{Metasurface functionality: Incident EM waves cause unit cells to
respond in a controlled manner for creating a specific EM response,
dictated by the selected state of active elements and the resulting
inductive surface currents. The cell-level engineered response can
be time-variant and follows a predetermined design goal such as directing
reflection towards a custom angle. The configuration DB, also known
as codebook, defines a format for a whole-tile configuration entries,
$\left[V\right]$, valid for a given tile design, and specific entries,
each corresponding to an EM response/function, $\left[v\right]$.}\label{fig:MSprinciple}
\end{figure*}

\paragraph*{Operating Principle}

In general, a metasurface operates by controlling surface currents
across its structure created via induction from impinging waves, as
shown in Fig~\ref{fig:MSprinciple}~\cite{ref12}. The
total electromagnetic response is calculated based on the emitted
field resulting from these currents. Engineering of surface current
dynamics involves considering direct, inductive currents on the passive
parts of the cells, currents wirelessly induced by neighboring cells,
and the current flow controlled by the tunable impedance circuit elements.
This can be conceptualized as a network of input and output antennas
connected through custom topologies using switch elements, where incident
waves are routed according to switch element states before exiting.
A departing wave of any specific form has the equivalent surface current
distribution that creates it, which in turn can manifest by transforming
the inductive waves via the circuit elements. 

This workflow, as shown in Fig~\ref{fig:MSprinciple}, can also be
a time-variant process. The metasurface can change its response while
the same wave impinges upon it, e.g., in order to steer it to another
direction, or when the impinging wave changes, in order to adapt to
it. Moreover, the time-scale of this time-variant behavior can be
larger than $\lambda$ or smaller than it. In the latter case, the
impinging signal remains intact and is redirected to another direction,
or is absorbed. In the former case, the metasurface \emph{alters}
the impinging signal. For instance, consider a single sinusoid of
wavelength $\lambda$ as the impinging wave. If the metasurface alternates
between steering and absorption per $\nicefrac{\lambda}{2}$, the
response would be an ON-OFF keying-modulated steered signal.
\begin{figure}[!t]
\begin{centering}
\includegraphics[width=1\columnwidth]{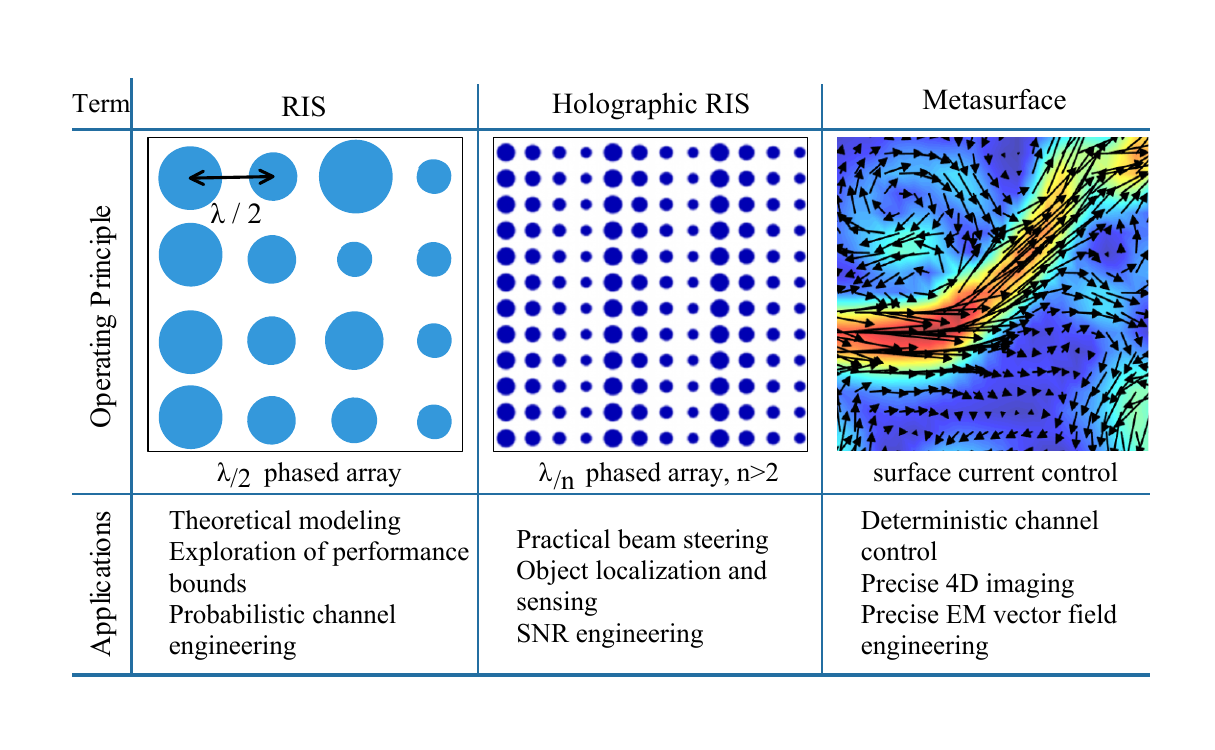}
\par\end{centering}
\caption{Overview of the usual metasurface-related nomenclature and their
main applications in the literature.}\label{fig:MStypes}
\end{figure}

\paragraph*{Nomenclature}

This general operating principle of metasurfaces as manipulators of
inductive surface currents is commonly simplified, exchanging performance
for ease of modeling and/or manufacturing. In particular, metasurfaces
are oftentimes treated as synonymous to antenna arrays, where each
cell is an isolated or coupled phase shifter, disallowing the direct
flow of inductive current between neighboring cells~\cite{ref13,ref14}.
This electric isolation is also enforced in the z-axis, where phase
shifter arrays are stacked for better performance, while metamaterials--the
3D counterpart of metasurfaces--remain fully electrically connected
in the vertical direction as well. These simplifications are typically
introduced to enable analytically tractable channel models and formulations,
rather than to faithfully represent the full extend of the electromagnetic
capabilities of metasurfaces.

This simplification has often appeared under multiple--and often
changing--naming conventions in the literature, such as reconfigurable
intelligent surfaces (RIS) and holographic RIS (HRIS), denoting sparser-to-denser
phased arrays, as shown in Fig.~\ref{fig:MStypes}. Nonetheless,
it is clarified that such simplifications result in a non-negligible
loss of control over the electromagnetic propagation compared to actual
metasurfaces, such as the inability to achieve steep angles in steering,
narrow-band operation, and ripples in reflection patterns that give
rise to deep fading phenomena in cascading reflection setups, which
are not limitations of metasurfaces in general~\cite{ref2}.
Therefore, RIS-based abstractions should be understood as intentional
reductions of the available wave control capabilities, rather than
as physically equivalent realizations of metasurfaces.This is also
reflected in their most prominent application orientation, summarized
in Fig.~\ref{fig:MStypes}. 

Certain studies, particularly on RIS, have also introduced a system
control models that are intended mainly for theoretical analysis.
In particular, while seeking the performance bounds of a RIS unit,
studies consider an optimization problem where, e.g., the SNR maximization
is the driving objective and every single phase shifter of every RIS
is an optimization variable~\cite{ref13,ref14}.
Given that a single RIS can have thousands of phase shifters, it becomes
apparent that this control approach is not intended for practical
application. In contrast, real time control of multiple metasurfaces
seeks to optimize macroscopic system level variables, such as reflection
and scattering directions, the distribution of power among outgoing
waves, and absorption behavior. The states of the embedded switches
that, for example, manifest the steering from a direction of arrival
to a reflection direction, are drawn from a database--known as \emph{codebook}--populated
during the manufacturing of a metasurface, as shown in Fig.~\ref{fig:MSprinciple}. 

Actual metasurfaces have had exciting applications initially in the
optical regime, such as micro-lenses~\cite{ref15} and programmable optical
scatterers~\cite{ref16}, which led to applications in fields such as
optics, spectroscopy, and high-resolution, compact imaging~\cite{ref17}.
In addition to their optical applications, metasurfaces also exhibit
promising properties in controlling electromagnetic fields~\cite{ref18}.
For example, metasurfaces can be designed to absorb or scatter electromagnetic
radiation, making them suitable for applications such as radiative
cooling and electromagnetic shielding. Furthermore, metasurfaces have
been used to enhance the efficiency of solar cells by manipulating
the interaction between incident light and the photovoltaic material. 

In the following, we focus on applications to wireless communications
as the driving scenario of the tutorial. The same modus operandi,
however, remains applicable for other cases, such as RF imaging. Moreover,
we will assume actual metasurfaces, i.e., without loss of modeling
generality. It is noted that the inductive current manipulator model
encompasses the phased array derivations of Fig.~\ref{fig:MStypes}
and, as such, the following remains applicable in their case.

\subsection{Manufacturing Approaches for Electromagnetic Metasurfaces}

Electromagnetic metasurfaces constitute the focus of the present paper,
as they have recently received the research and industrial spotlight
in the context of 6G integrated sensing and communications~\cite{ref19}.
Their fabrication poses significant challenges due to the need for
precise control over the periodic structure's geometry, material properties,
and interconnects. In the following, we survey some of the most significant
manufacturing approaches from two aspects: 
\begin{enumerate}
\item The physical aspect, encompassing the passive materials (e.g., plain
metal parts) and active parts that are involved in the electromagnetic
response of a metasurface. In the following, we will describe the
active parts as dynamic impedance-shifting components, which constitute
a generalization over phase-shifters, thus covering both metasurfaces
and antenna arrays.
\item The logic control aspect, involving the logic circuitry required for
the interaction between a computer and the metasurface, in order to
modify its electromagnetic response and overall behavior. This aspect
determines how metasurfaces can be integrated into software controlled
networks and how rapidly and reliably configuration decisions can
be enforced.
\end{enumerate}
\textbf{The physical aspect}. This aspect comprises all parts of the
metasurface that allow impinging waves to form inductive currents
within or over them, as well as materials that provide structure integrity
and electromagnetic isolation. Corresponding examples include copper
patches, PIN diodes, dielectric substrates and prepreg materials~\cite{ref20}.
Major manufacturing approaches and options are the following:

\emph{Lithography and Etching}. Lithography and etching techniques
are widely used for fabricating electromagnetic metasurfaces~\cite{ref21}.
These methods involve pattern transfer onto a substrate using photoresist
materials, followed by etching or removal of material to create the
desired microstructure. The resulting structures can be used to create
periodic arrays with precise control over geometry and material properties.

\emph{Nanoimprint lithography} (NIL) is another approach that has
been explored for fabricating electromagnetic metasurfaces~\cite{ref22}.
This method involves creating a master structure with nano-scale features,
which are then transferred onto a substrate using NIL in a stamping
manner. The resulting structures can be used to create periodic arrays
with enhanced electromagnetic properties.

\emph{3D Printing}. Three-dimensional printing (3DP) has also been
explored for the fabrication of electromagnetic metasurfaces~\cite{ref23}.
This approach involves creating a 3D structure with nested layers
of material that are precisely controlled in terms of geometry, density,
and composition. The resulting structures can be used to create complex
periodic patterns that exhibit enhanced electromagnetic properties,
while from a system perspective, such flexibility allows the exploration
of non conventional metasurface geometries, potentially enabling novel
control functions.

The following commonly come as packaged electronic components, that
are employed as tunability agents within the metasurface. However,
as they constitute parts that allow the flow of EM inductive currents,
they are studied as part of the physical aspect of a metasurface.

\emph{Varactor-based Impedance Shifters}. Varactor diodes are widely
used for impedance shifting due to their simplicity and tunability~\cite{ref24}.
By varying the capacitance of the varactor, it is possible to achieve
a controlled impedance shift between incoming and outgoing waves. 

\emph{MEMS} are employed in some impedance shifting applications due
to their high precision and low power consumption~\cite{ref25}. By
physically altering the path length of the wave, MEMS can achieve
precise control over impedance shift. This technique is commonly found
in advanced applications requiring high performance.

\emph{Liquid crystals} are another option for impedance shifting,
particularly useful for tunability across a wide frequency range~\cite{ref26}.
The refractive index of liquid crystals can be controlled by an electric
field, enabling rapid phase modulation. This technique offers flexibility
and precision in metasurface design.

\emph{Graphene-based impedance Shifters}. Graphene's electrical tunability
makes it a promising material for impedance shifting~\cite{ref27}.
By controlling the voltage bias, graphene's properties can be altered
to achieve desired impedance shifts. Graphene-based impedance shifters
are an area of active research due to their remarkable electronic
properties.

\emph{Ferroelectric impedance Shifters}. Materials with ferroelectric
properties are utilized for impedance modulation due to their high-speed
switching capabilities~\cite{ref28}. Ferroelectric impedance shifters
offer rapid and efficient control over impedance shift, making them
suitable for applications requiring fast response times.

\emph{Mechanically Reconfigurable impedance Shifters}. Mechanical
changes can be used to alter the impedance configuration in metasurfaces~\cite{ref29}.
While this technique is simpler than others, it is useful for slower-adapting
application scenaria. Mechanically reconfigurable impedance shifters
offer flexibility and low power consumption, making them suitable
for specific use cases. 
\begin{table*}
\centering{}\caption{Comparison of Impedance Shifter Types.}\label{table:phase_shifters}
\begin{tabular*}{1\textwidth}{@{\extracolsep{\fill}}@{\extracolsep{\fill}}|c||>{\centering}m{30mm}|>{\centering}m{30mm}|>{\centering}m{30mm}|>{\centering}m{30mm}|}
\hline 
\multirow{1}{30mm}{\textbf{Impedance Shifter Type}} & \textbf{Efficiency } & \textbf{Bandwidth } & \textbf{Manufacturing Cost } & \textbf{Energy Consumption}\tabularnewline
\hline 
\hline 
Varactor-based  & Moderate  & Moderate  & Low  & Low\tabularnewline
\hline 
MEMS-based  & High  & Wide  & Moderate to High  & Minimal (state-preserving)\tabularnewline
\hline 
Liquid Crystal  & Moderate  & Wide  & Moderate  & Low to Moderate\tabularnewline
\hline 
Graphene-based  & High  & Potentially Wide  & High  & Low\tabularnewline
\hline 
Ferroelectric  & High  & Moderate to Wide  & Moderate  & Low to Moderate\tabularnewline
\hline 
Mechanically Reconfigurable  & Moderate  & Moderate  & Low  & Minimal (state-preserving)\tabularnewline
\hline 
\end{tabular*}
\end{table*}

A qualitative comparative of these options are given in Table~\ref{table:phase_shifters}.
In terms of efficiency, high degree of control over the current flow
is typically observed in impedance shifters that utilize precise impedance
control mechanisms, such as MEMS, graphene, ferroelectric, metamaterial-based,
and photonic solutions. These technologies enable the achievement
of high-quality impedance shifting with minimal losses, making them
suitable for a wide range of applications. In terms of bandwidth,
MEMS and liquid crystal-based impedance shifters exhibit very wide
bandwidth, which is essential for advanced applications that require
precise control over the impedance-shifted signal. This high bandwidth
enables these solutions to operate effectively in environments where,
e.g., rapid phase modulation is necessary. The manufacturing cost
of these options varies widely depending on the technology employed.
Graphene and photonic shifters are generally more expensive due to
the advanced materials and processing techniques required, while simpler
solutions such as varactor and mechanically reconfigurable types are
less costly. 

In terms of energy consumption, low consumption is a primary objective
for impedance shifters, as it is essential for their deployment in
practical applications. Most impedance shifter solutions aim to minimize
energy consumption while maintaining high performance. MEMS and graphene-based
solutions are particularly noteworthy in this regard, as they exhibit
low power requirements that make them well-suited for efficient operation.
However, electronic impedance shifters usually require power supply
to maintain a certain impedance shift state. MEMS-based and mechanical
shifters may have: i) lower tuning speed, and ii) higher initial state
shift consumption than their electronics counterparts, but are \emph{state-preserving},
i.e., they do not require any power to maintain their set state. Thus,
their energy consumption is marked as minimal in Table~\ref{table:phase_shifters}.
Notice that this depends on the impedance shifting frequency, since
in fast shifting cases, the overall consumption may be defined mainly
by the initial state shift energy expenditure. In other words, this
distinction becomes important when metasurface reconfiguration is
performed frequently or under strict energy budgets.

\paragraph*{On assembly prospects}

It is important to note that each of the aforementioned types of impedance
shifters may require different assembly approaches. Since Printed
Circuit Board (PCB) processes are usually more accessible to early
research prototyping efforts~\cite{ref30}, a
possible misconception is that metasurfaces are required to come in
PCB form. However, as large-scale industrial and real-work deployments
draw closer, other assembly processes become more fitting. In particular,
Large Area Electronics (LAE) systems involve the direct fabrication
of functional electronics on expansive substrates, such as flexible
or three-dimensional materials, through innovative manufacturing techniques
like inkjet printing, screen printing, and evaporative deposition~\cite{ref31}.
This approach enables seamless integration of electronic components
into everyday items such as smart clothing, autonomous vehicles, and
intelligent structures, unlike conventional PCB assemblies that are
typically confined to flat, solid substrates with form-factor and
overall size limitations. By employing innovative fabrication methods
such as hybrid CMOS/LAE architectures and programmable micro-assemblies,
LAE systems have become known for their design flexibility, functionality
and scalability~\cite{ref32}. In contrast to traditional
surface mount technology (SMT)/through-hole technology (THT) PCB assembly
processes, which are limited by their reliance on conventional materials
and manufacturing techniques, LAE enables in-situ integration of advanced
materials and components, making them excellent candidates for massive
metasurface manufacturing and deployment~\cite{ref20}.

\textbf{The logic control aspect}. The control logic circuitry of
metasurfaces enables the scalable and precise manipulation of electromagnetic
waves based on external software commands, effectively forming the
interface between network level control decisions and the realization
of electromagnetic functionalities. The choice of logic control circuitry
depends on the specific requirements of the application, including
the need for flexibility, reconfigurability, and performance.

One popular approach is the use of \emph{Field-Programmable Gate Arrays
(FPGAs)}, which offer flexible and reconfigurable logic that can be
easily adapted to changing needs~\cite{ref33}. FPGAs are well-suited
for prototyping and dynamic system configuration, while commonly supporting
a very large number of control pins per single package, a trait that
matches well with the numerous control elements typically embedded
in a metasurface. Additionally, their ability to handle complex logic
operations ensures efficient processing of control signals~\cite{ref34}.

For simpler systems, \emph{microcontrollers} provide an efficient
and cost-effective solution for managing control logic~\cite{ref35}.
These small computers can effectively control individual elements
of a metasurface, offering a straightforward approach to system management.
However, as the complexity of the system increases, more advanced
logic control circuitry may be required.

\emph{Application-Specific Integrated Circuits (ASICs)} constitute
a tailored solution in cases of specific speed, energy efficiency
and/or footprint requirements~\cite{ref36}. Optimized for specific
tasks, ASICs provide the best possible performance but come with higher
development costs and limited flexibility.

\emph{Digital Signal Processors} (DSPs) are an option for metasurfaces
requiring advanced signal manipulation capabilities, e.g., to remodulate
the inductive currents in specific ways~\cite{ref37}. DSPs can execute
complex algorithms efficiently, enabling dynamic tunability and adaptive
systems. Such processing capabilities may become relevant in cases
where the metasurface behavior must occur in response to specific
attributes of the incoming wave, i.e., at a complex waveform-processing
level.

Another approach is to integrate both analog and digital components
into \emph{Mixed-Signal Circuits}~\cite{ref38}, which can handle the
analog nature of wave manipulation in metasurfaces while maintaining
digital control for programmability. This hybrid approach enables
the efficient processing of control signals while preserving the flexibility
required for system integration and adaptation. In this sense, mixed
signal designs offer a natural continuation of purely digital control,
allowing part of the signal processing burden to be handled closer
to the physical interaction with the wave.

Finally, \emph{System-on-Chip (SoC) solutions} offer a comprehensive
integration of multiple components, including processors, memory,
and control logic, into a single chip~\cite{ref39}. SoC architectures
can be designed to include specialized modules for metasurface control,
providing both high performance and integration. By leveraging these
advanced logic control circuitry options, researchers and engineers
can develop efficient and adaptable metasurfaces that optimize electromagnetic
wave manipulation for a wide range of applications. This increasing
degree of integration becomes particularly relevant as metasurfaces
scale in size and complexity, since it directly affects footprint,
reliability, and deployment practicality.
\begin{table*}
\centering{}\caption{Logic Control Circuitry Options for Metasurfaces}\label{tab:logic_control_circuitry}
\begin{tabular*}{1\textwidth}{@{\extracolsep{\fill}}@{\extracolsep{\fill}}|l|l|l|l|l|l|l|}
\hline 
Aspect  & FPGAs  & Microcontrollers  & ASICs  & DSPs  & Mixed-Signal Circuits  & SoC Solutions \tabularnewline
\hline 
Flexibility  & High  & Moderate  & Low  & Moderate  & Moderate  & High \tabularnewline
Performance  & High  & Moderate  & Very High  & High  & High  & High \tabularnewline
Power Consumption  & Moderate to High  & Low  & Very Low  & Moderate  & Moderate  & Low to Moderate \tabularnewline
Development Cost  & Moderate to High  & Low  & Very High  & Moderate  & High  & Moderate to High \tabularnewline
Complexity of Design  & High  & Low to Moderate  & High  & Moderate  & Moderate  & High \tabularnewline
Speed of Operation  & Very High  & Low to Moderate  & Very High  & High  & Moderate  & High \tabularnewline
Real-Time Processing  & Excellent  & Limited  & Not applicable  & Excellent  & Limited  & Excellent \tabularnewline
Reconfigurability  & Excellent  & Limited  & None  & Limited  & Limited  & Good \tabularnewline
Suitability for Prototyping  & Excellent  & Good  & Poor  & Moderate  & Moderate  & Good \tabularnewline
\hline 
\end{tabular*}
\end{table*}

The choice of the logic control also affects the wiring within the
metasurface, which usually comes in the form of patterned metallic
traces. For instance, FPGAs can support many physical pinouts, each
directly controlling an active element. Moreover, they can implement
assistive tasks, such as logic control break-out, by implementing
impedance shifters and in some cases DC voltage regulators. It is
also expected that a specifically programmed FPGA also provided very
low control latency.

Employing simple microprocessors instead of FPGAs, has the benefit
of lower cost and easier development, but can require external assistive
circuitry that must be placed upon the metasurface, yielding extra
footprint and wiring. Towards cost-effectiveness, meta-atoms can also
be controlled via shared means, such as buses and cross-bars, reducing
the wiring complexity and costs further, but usually the expense of
control latency. This trade off illustrates how hardware control choices
influence the temporal granularity at which metasurface configurations
can be updated. Notably, there also exist theoretical proposals for
metasurfaces comprising wireless cells, both for configuration and
for energy harvesting~\cite{ref40}.

\section{Metasurfaces as Network Components}\label{sec:Metasurfaces-as-Network}

Having covered basic prerequisites from the physical layer of metasurfaces,
we proceed to study their network-level modeling. In the following,
we will seek to align with the concepts of network graphs, network
routing and virtual network functions, which are basic working tools
in the field of networking~\cite{ref41,ref42,ref43}. 
\begin{rem}
The ensuing model of metasurfaces as network components does not imply
a store-and-forward operation model, given that metasurfaces cannot
be aware of the data carried by impinging waves.
\end{rem}
Instead, the following model is aligned to the pass-through network
switching, which is used extensively in optical networks~\cite{ref44}.

\subsection{Modeling Metasurfaces as Wave Routers}\label{subsec:collimate}

As discussed, metasurfaces provide advanced electromagnetic control
and high granularity capabilities for various applications like wide-band
communications, energy harvesting, ultra-high resolution medical imaging,
sensing, quantum optics, and military use cases~\cite{ref45}.
They are well known for their ability to create pencil-like beams
(known as collimation), non-reciprocating behavior (i.e., reflecting
towards one direction with excellent efficiency, but absorbing in
the reverse), as well as operating as frequency filters. Moreover,
designs exist for virtually any operating band, from MHz to THz and
beyond. 
\begin{figure}[!t]
\begin{centering}
\includegraphics[width=1\columnwidth]{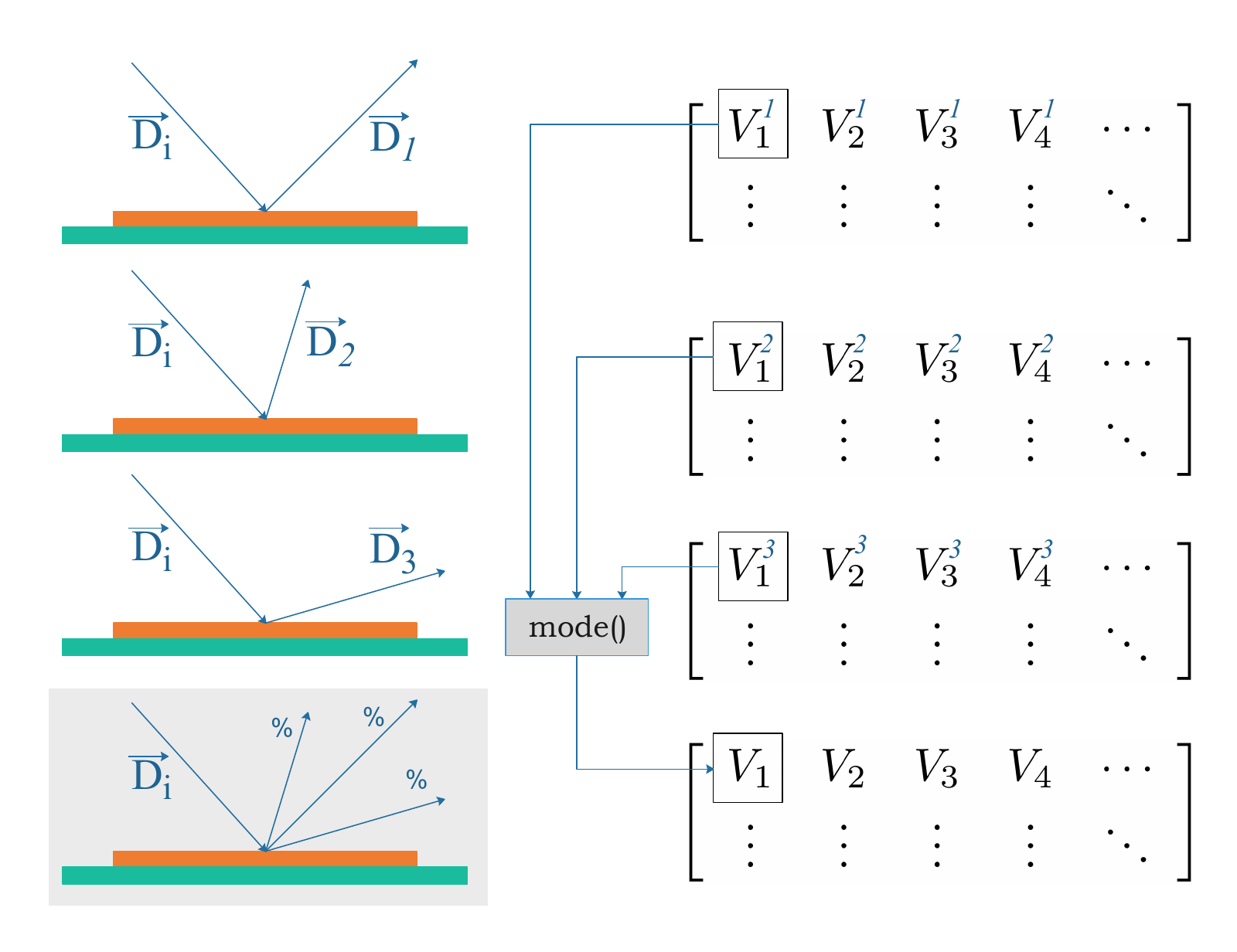}
\par\end{centering}
\caption{A process of compiling complex EM metasurface Functions from basic
ones~\cite{ref46}. In the illustrated example, a 3-way
split is compiled from three basic wave steer EM Functions. (The $\texttt{mode}$
is the value that appears most often in a set of data values, following
a rounding step).}\label{fig:merge}
\end{figure}

Metasurfaces can support multiple EM interactions including reflection,
refraction, absorption, and polarization. These four interactions
can be considered as basic \emph{EM Functions}, as shown in Fig.~\ref{fig:MSprinciple}.
Complex functionalities can be built by combining these basic functions,
as shown in Fig.~\ref{fig:merge}. This combining occurs by following
merging rule on the corresponding configurations of the EM Functions
to be merged. For instance, a new, complex EM Function can be created
by defining $V_{i}^{*}$ as the $\text{\texttt{mode}}{}_{\forall i}\left\{ V_{i}\right\} $,
$V_{i}$ being the bias signal value applied to the $i^{th}$ cell,
as shown in Fig.~\ref{fig:MSprinciple} and~\ref{fig:merge}. Naturally,
merging multiple functionalities results into a lower degree of efficiency
for each one, compared to the case when only a single EM Function
is activated. 

For instance, let $f_{1}$ denote a specific EM Function performing,
e.g., steering of waves incoming from a direction $\overrightarrow{i_{1}}$,
towards a direction $\overrightarrow{o_{2}}$. Let $e_{1}^{i}$ be
the total power impinging upon the corresponding metasurface from
the intended direction $\overrightarrow{i_{1}}$, and $e_{1}^{o}$be
the total power that the metasurface manages to reflect towards $\overrightarrow{o_{2}}$.
The efficiency $\epsilon_{1}$ of $f_{1}$ is defined as~\cite{ref47}:
\begin{equation}
\epsilon_{1}=\frac{e_{1}^{i}}{e_{1}^{o}}.\label{eq:epsilon}
\end{equation}
Proceeding, consider a second EM Function, $f_{2}$, which can freely
perform a different type of steering, direction-dependent absorption,
etc. Let $\epsilon_{2}$ be its efficiency when being deployed alone
in the same metasurface. Now lets consider that both $f_{1}$ and
$f_{2}$ are deployed together in the metasurface by following the
$\texttt{mode}$ rule mentioned above. 
\begin{rem}
The $\texttt{mode}$ rule yields a new combined functionality performing
both tasks at the same time. However, when measured separately, the
updated efficiency of each function will be: 
\begin{equation}
\epsilon_{1}'\le\epsilon_{1}\,\text{and}\,\epsilon_{2}'\le\epsilon_{2},\label{eq:dropEffic}
\end{equation}
i.e., the efficiency of each function will be reduced. The equality
in relation (\ref{eq:dropEffic}) holds on special cases where two
or more EM functions can be served with the same $\left\{ V_{i}\right\} $
values. 
\end{rem}
Notice that one can freely combine any number of function minding,
however, the associated efficiency drop. Further details on this topic
can be found in the related literature, covering the cases of how
to combine multiple EM functions to attain some specific efficiency
degrees, and how to price them using a resource sharing paradigm~\cite{ref47}.
\begin{figure}[!t]
\centering{}\includegraphics[width=1\columnwidth]{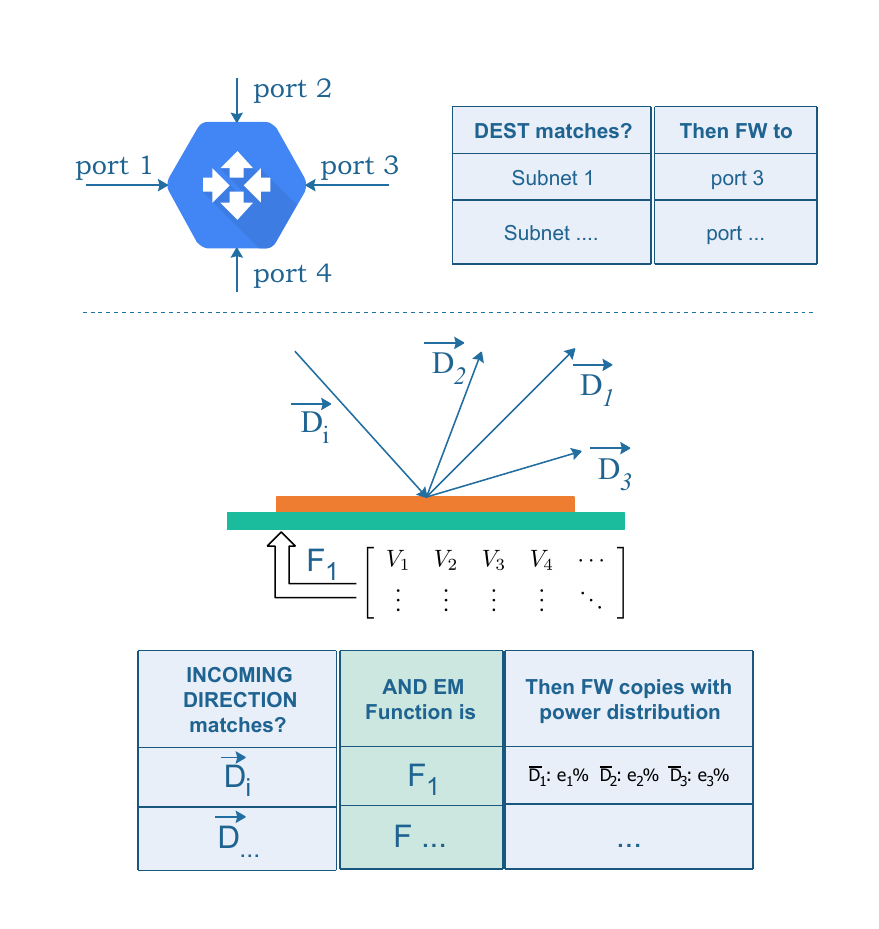}\caption{Visualization of the router model for metasurfaces. Top: the classic,
destination-based routing process found in wired networks, following
a port forwarding table. Bottom: an equivalent table form, describing
the incoming wave redirection achieved by a metasurface. The redirection
outcome depends on the incoming direction and the activated EM Function.}\label{fig:graph}
\end{figure}

The described concept of EM Functions can be leveraged when configuring
a single or a set of metasurfaces to serve the objective of user(s),
which cannot consider the direct optimization of the active elements
due to practical reasons. A single metasurface can have thousands
of active elements, each with a multitude of possible values. Thus,
treating them as direct optimization variables is impractical, especially
for real-time operation.

Instead, metasurfaces can be modeled in terms of macroscopic inputs
and outputs, i.e., directions of arrival and departure expressed via
the concept of EM Functions. Optimizing the behavior of a metasurface
then corresponds to an objective of connecting input directions to
output directions, which has the benefits of: 
\begin{itemize}
\item Reducing the dimensionality of the problem, compared to the direct
optimization of impedance shifts. 
\item Allowing for a model akin to that of routers in classic wired networks~\cite{ref8},
which is very well-known in the networking field, and comes with a
framework for ensuring connectivity, following graph path-finding
approaches. 
\end{itemize}
Thus, the proposed model integrates a forwarding table into each metasurface,
which is analogous to the concept of forwarding tables found in computer
networks, and is visualized in Fig.~\ref{fig:graph}. This integration
serves several purposes:
\begin{itemize}
\item \emph{Treat wave directions as discrete input/output ports via Column
1}. In essence, not every possible direction vector is practically
useful for the operation of the system as a whole. A metasurface seeks
to reflectively connect users and metasurfaces, i.e., form metasurface-metasurface,
user-metasurface or user-user reflections. From these discrete targets,
metasurfaces have fixed relative points in space, while the position
of users can also be discretized. 
\item \emph{Describe the metasurface behavior under different EM functions
via Column 2}. This entry also lists the supported EM functions, i.e.,
configurations that have been tested and profiled during the manufacturing
stage. 
\item \emph{Describe the scattering radiation as a ``leaky multicast''
action via Column 3}. This entry includes profiling measurements pertaining
to an EM Function identifier and an incoming wave direction. Following
the aforementioned profiling process by the manufacturer, it consists
of a hash map connecting output directions/ports to the associated
power loss measured exactly when departing the metasurface. Thus,
the scattering phenomenon becomes akin to packet multicasting, albeit
with the extra attribute of associated power loss. (A path loss rule
can yield the total power losses at a distance along an output port). 
\end{itemize}
Thus, by combining a forwarding table functionality with EM Function
support and profiling measurements, the proposed router model for
metasurfaces offers a connection to existing framework for managing
and configuring complex networks, even in real-time. 
\begin{rem}
The routing table may not be stored as an exhaustive list of entries.
In general, it is a service that produces a resulting power distribution,
given an incoming direction of a selected EM Function. 
\end{rem}
In the following, we leverage the router model to treat a complete
metasurface-enabled wireless system as a graph, and formulate configuration
objectives as path finding problems.

\subsection{Describing Systems of Metasurfaces}

In everyday communication scenarios, multiple wireless users often
interact within physical spaces. In such environments, electromagnetic
waves freely dissipate, leading to interference and potential eavesdropping
among devices.

In contrast, the concept of PWEs employs metasurface coatings placed
on walls and ceilings in the form of tile-like units, in order: i)
to describe systems of metasurfaces, and ii) offer user-defined wireless
propagation~\cite{ref1}. PWEs receive software commands
form network administrators and adapt their interaction with electromagnetic
waves to meet user needs. For instance, Figure~\ref{fig:nnpaths}
(top) demonstrates a situation where a user requires enhanced security
against eavesdropping. The programmable environment collaborates with
user devices to establish an \textquotedbl air-path\textquotedbl{}
that avoids the living space of other users, thereby minimizing interference
and the prospects of eavesdropping. 

Conversely, other users that do not require such a level of privacy
guarantee, can be handled by a global wireless policy which, e.g.,
optimizes their data transfer rates by reducing cross-interference
and fine-tuning their wireless reception profile. Other service examples
include the remote charging of inactive wireless devices and, should
a user fail to meet the network's access policies, an environment-activated
blocking or absorption of his emissions, potentially harnessing the
energy for constructive use.

We proceed to describe a unified way for modeling, configuring and
interfacing PWEs using the concept of graphs~\cite{ref8}. 

\subsection{A Graph Model for Networked Metasurfaces}\label{subsec:A-Graph-Model}

\begin{figure}
\begin{centering}
\includegraphics[width=1\columnwidth]{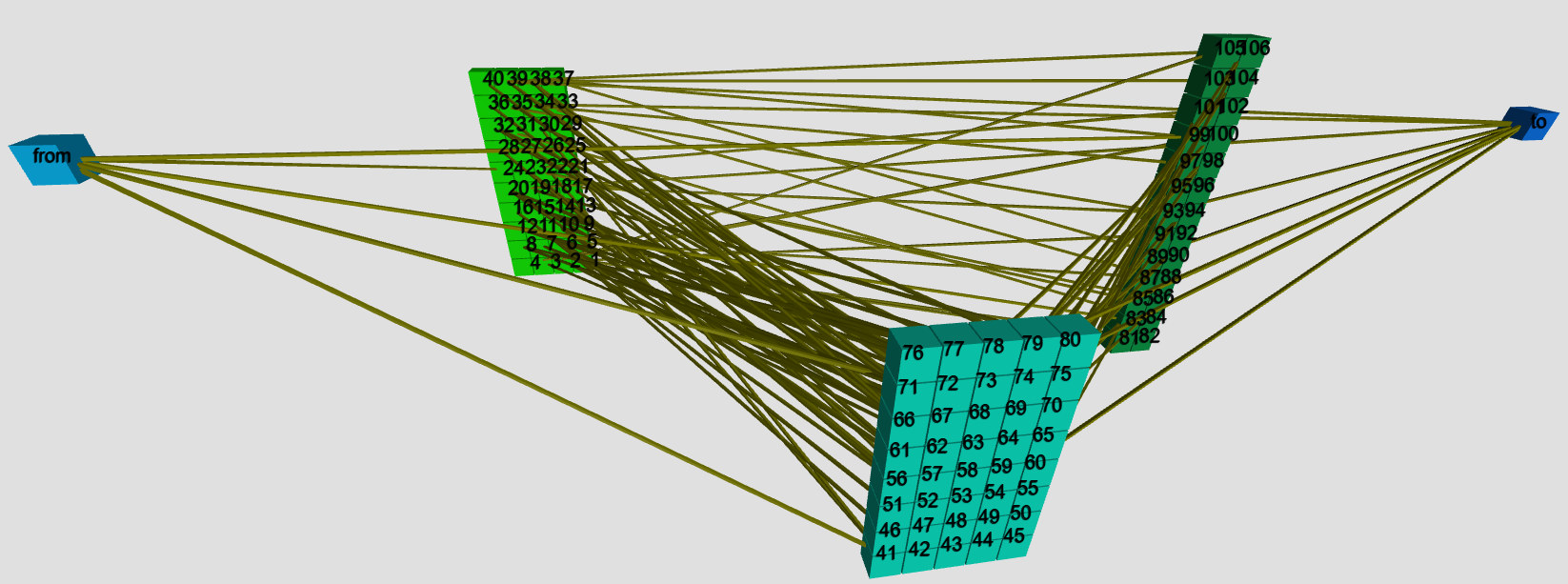}
\par\end{centering}
\centering{}\caption{An example of a PWE graph resulting from a 3D arrangement of metasurface
tiles over wall segments.}\label{fig:graphs}
\end{figure}
When signals travel through a three-dimensional space, they can take
two main paths: Line-of-Sight (LOS), where the signal travels directly
from transmitter to receiver without obstruction, and Non-LOS (NLOS),
where the signal is scattered or reflected before reaching its destination.
PWEs primarily influence the NLOS component, leaving LOS untouched.
Our focus will be on understanding NLOS scenarios within PWEs, while
accounting for the existence of LOS waves.

Imagine two metasurface tiles in a 3D space. These tiles are considered
connectable if:
\begin{itemize}
\item There's an input signal from the environment.
\item Each tile has at least one EM Function that allows them to interact
with this signal.
\end{itemize}
In simple terms, two tiles are connectable if one can redirect incoming
waves to the other, given that this redirected power is significant
enough for the application in question. However, physical barriers
between tiles or a lack of appropriate tile functions can hinder this
connectivity.

In the same space, consider a set of user devices. A user device is
considered connected to a tile if:
\begin{itemize}
\item There's a direct LOS input signal from the user to the tile.
\item Zero reflection from the tile is achievable only via an EM Function
dictating full absorption.
\end{itemize}
In this context, we define some connectivity links as follows: i)
Inter-tile links,which represent potential connectivity between tiles,
as defined above, and ii) user-tile links, which indicate connections
between user devices and tiles, based on the outlined conditions.

A key characteristic of these links is that they are bidirectional
and symmetric, i.e., they affect the signal exactly the same in both
directions. Additionally, each link inserts a propagation delay to
any wave that traverses it. This delay stems from the total length
of the link, considering a constant EM propagation speed within the
given space. Moreover, and in order to capture Multiple-Input-Multiple-Output
(MIMO) capabilities, links may be labeled to identify transmitting
or receiving user devices and antennas.

With these inputs, we define the graph representation of a PWE, comprising
the sets of tiles, user devices, inter-tile links and user-tile links,
with an example given in Fig.~\ref{fig:graphs}. Subgraphs, formed
by applying specific EM functions per tile, give rise to the potential
paths within this environment: Consider a wave flow from a transmitter
to a receiver, navigating through tiles in a configured PWE, i.e.,
a PWE with EM Functions activated in its tiles. This path is an ordered
selection of links without repetition, illustrating how PWEs facilitate
or hinder communication based on their setup.

Here we make the following remarks:
\begin{rem}
The graph model can cover cases of partial metasurface coverage in
a space (such as absence of metasurfaces over the floors). In order
to achieve this, the non-coated surfaces are assumed to be covered
with virtual metasurface tiles that support only one EM Function,
i.e., that of specular/natural (and potentially lossy, as dictated
by the corresponding materials) reflection and diffraction.
\end{rem}
Moreover:
\begin{rem}
\label{rem:There-exist-visibility}There exist visibility cases that
fall between LOS and NLOS, known as near-LOS or nLOS, which attenuate
the signal more than plain LOS. The corresponding links in the PWE
graph model carry an extra attribute that expresses this constant
power attenuation in the form of a multiplicative factor.
\end{rem}
It is noted that nLOS is quantified by the theory of Fresnel zones,
and more information can be found in the related literature~\cite{ref48}. 

\section{PWE Performance Optimization and Objective Modeling}\label{sec:PWE-Performance-Optimization}

Given that PWEs represented a wireless propagation setting as a graph,
the problem of optimizing the EM Function for each metasurface tile
can be formulated as a graph-theoretic problem. The general optimization
objective is to engineer a critical channel quantity, known as Power-Delay
Profile (PDP), by routing the waves emitted by the Tx within the PWE
graph.

In the context of wireless communication systems, PDP refers to the
spatial distribution of signal power versus arrival delay at a receiver
location within a wireless channel. The PDP characterizes the behavior
of the transmitted wave as it propagates through the channel, taking
into account various factors that affect the signal's amplitude and
timing~\cite{ref49}.

Mathematically, the PDP can be represented as a two-dimensional function,
where each point in the function, indexed as $k=1\ldots K$, corresponds
to a specific route that the wave undertook in order to eventually
reach the receiver. Each point is marked by an arrival delay ($\tau_{k}$)
measured from the time the signal was transmitted, and a received
power level ($P_{k}$). In essence, the PDP describes how the received
copies of the original signal (due to reflection, refraction, etc.)
get distributed over time and power, providing insight into the channel's
characteristics and behavior. Understanding the PDP is crucial in
wireless communication systems, as it informs the design of transmission
strategies, modulation schemes, while defining the achievable data
transfer rates.

PWEs can craft the PDP per user pair by routing a signal via the PWE
graph and over specific paths with known delay, and potentially predictable
power loss~\cite{ref8}. This engineered PDP could lead
to the mitigation of multipath fading phenomena. Alternatively, a
\textquotedbl multipath amplifier\textquotedbl{} effect can be enforced
when signal obfuscation is required instead. PWEs can create multiple
simultaneous signals with destructively superposing arrival times
and amplitudes, simulating complex multi-path environments. In general,
by precisely controlling the wave routing process, PWEs can engineer
the PDP of the wireless channel to test various transmission strategies,
modulation schemes.

While the PDP optimization constitutes an overarching goal, it is
possible to define simpler metrics based on it, in order to facilitate
the optimization computations. One such approach is to define a simple
scalar metric over the PDP data. One such metric, commonly used in
wireless communications, is the Root Mean Squared-Delay Spread (RMS-DS),
defined as~\cite{ref49}:
\begin{equation}
R_{\sigma}=\sqrt{\overline{\tau^{2}}-\left(\bar{\tau}\right)^{2}}
\end{equation}
where
\begin{equation}
\bar{\tau}=\frac{\sum_{\forall k}P_{k}\cdot D_{k}}{\sum_{\forall k}P_{k}}\,\text{ and }\,\overline{\tau^{2}}=\frac{\sum_{\forall k}P_{k}\cdot D_{k}^{2}}{\sum_{\forall k}P_{k}}.
\end{equation}
A channel model with low $R_{\sigma}$ corresponds to a multipath
setting where the PDP is closer to equalization, i.e., minimal fading
effects. However, a deceptively minimal $R_{\sigma}$ value can also
stem from keeping just one path out of the $\left\{ P_{k},\tau_{k}\right\} $
set, while the rest are driven away from the receiver, hence reducing
the overall received power. 

Thus, defining PWE optimization metrics over the PDP typically needs
to consider accompanying conditions that should be upheld from a PWE
configuration solution. 

\subsection{An Optimization Framework for PWEs}\label{subsec:A-VNE-Embedding}

Consider a PWE with a set of users $u\in U$, and a set of performance
metrics, $\mathbf{M}$, from which each user can select a subset $M_{u}\subseteq\mathbf{M}$,
iterated as $\mu_{u}\in M_{u}$, in order to define his requirements
from the PWE. Examples can include the maximization of his signal-to-interference-plus-noise
ratio, the maximization of power harvesting rate, latency minimization,
the minimization of the signal exposed to other users, etc.

The PWE is composed of a set of tiles $N$, and a set of EM functions,
$\mathbf{F}$, from which each tile $n\in N$ supports a certain subset
$F_{n}\subseteq\mathbf{F}$, iterated as $f_{n}\in F_{n}$, due to
manufacturing diversity or placement properties. Examples include
PWEs with mixed metasurface and RIS units, and tiles placed in open
spaces versus under severe blockage with corresponding wave steering
diversity potential, respectively. The set $\mathbf{F}$ is considered
to contain all possible merged functionalities as well (described
in the context of Fig.~\ref{fig:merge}). Deactivation of tile $n$
(i.e., yielding regular, non-metamaterial behavior) will be denoted
as $f_{n}=\emptyset$.

Furthermore, we define the following comparator:

\begin{equation}
c\left(f,f^{*}\right)=\left\{ \begin{array}{cc}
1 & \text{iff}\,f=f^{*}\\
0 & \text{iff}\,f\ne f^{*}
\end{array}\right..\label{eq:comparator}
\end{equation}
PWEs need to be re-configured in order to adapt to changing user requirements,
user mobility or time-variant blockage in the environment. Consider
a notation describing the optimal PWE configuration at a round $t-1$,
and at round $t$ as $f_{n}^{t-1},\forall n$ and $f_{n}^{t},\forall n$,
respectively. Then, the number of updates required to migrate from
$f_{n}^{t-1}$ to $f_{n}^{t}$ is:
\begin{equation}
R^{t}=\sum_{\forall n}\left(1-c\left(f_{n}^{t-1},f_{n}^{t}\right)\right)=\left\Vert N\right\Vert -\sum_{\forall n}c\left(f_{n}^{t-1},f_{n}^{t}\right)
\end{equation}
Notably, one should keep the number of updates minimal, throughout
the operation rounds of a PWE. In other words, follow a proactive
policy that ensures that no major increase should occur in $R^{t}$,
as this can correspond to extensive changes in the PWE configuration,
increasing control latency and potentially leading to unintended behavior
(as shown later, in Fig.~\ref{fig:setupMobility}). Moreover, the
number of unused (free) tiles at round $t$ is: 
\begin{equation}
S_{\emptyset}^{t}=\sum_{\forall n}c\left(f_{n}^{t},\emptyset\right)
\end{equation}
In general, one should keep $S_{\emptyset}^{t}$ maximal at every
$t$, in order to both increase the service capacity of the PWE (i.e.,
be ready to accommodate changes in the next round), and to be economic
in terms of energy consumption and administrative overhead. However,
keeping $R^{t}$ minimal throughput the operation rounds is a more
pronounced concern over $S_{\emptyset}^{t}$.

We proceed to formulate the general optimization objective of a PWE
at time $t$ as follows:
\begin{equation}
\begin{array}{cc}
\text{\text{opt}:} & _{\left\{ f_{n}^{t}\in F_{n},\forall n\right\} }\left\{ \mu_{u}^{t}\in M_{u},\,\forall u\right\} \\
\text{min-max:} & R^{t},\,\forall t\\
\text{max:} & S_{\emptyset}^{t}
\end{array},\label{eq:generalOpt}
\end{equation}
where: 
\begin{enumerate}
\item the first line expresses the selection of tile functions for the fulfillment
of user requirements are first priority,
\item the second line expresses the constraint over successive PWE updates,
and 
\item the third line the maximization of free tiles in the system only at
round $t$, denoting it as last priority.
\end{enumerate}
Notice that the formulation (\ref{eq:generalOpt}) considers the transmission
and receptions characteristics of the users to be predefined. In other
words, parameters such as the transmission power and user antenna
directivity are considered to be set. This can be generalized by considering
the user devices that also support a corresponding set of special
EM Functions (slightly abusing the term definition to cover user devices
apart from tiles), where an optimal choice must also be made. 

In the general formulation (\ref{eq:generalOpt}), a solution $\left\{ f_{n}^{t}\in F_{n},\forall n\right\} $
is assumed to be Pareto-dominant across all three aforementioned concerns.
This approach on itself can be restrictive, as a valid solution may
potentially not be possible. Thus, more relaxed expressions can can
obtained such as by employing pre-defined limits $R^{max}$ and $S_{\emptyset}^{min}$
such that:
\begin{equation}
\begin{array}{cc}
\text{\text{opt}:} & _{\left\{ f_{n}^{t}\in F_{n},\forall n\right\} }\left\{ \mu_{u}^{t}\in M_{u},\,\forall u\right\} \\
\text{subject to:} & R^{t}\le R^{max}\\
 & S_{\emptyset}^{t}\ge S_{\emptyset}^{min}
\end{array},\label{eq:generalOpt-1}
\end{equation}
where the latter two conditions should be treated as weak. Other simplifications
would be to incorporate the three clauses of formulation~(\ref{eq:generalOpt})
into one sum using weights, as per the standard practice for simplifying
Pareto-dominant formulations. We proceed to make the following further
remarks regarding: 
\begin{itemize}
\item the consistency of the updates, and
\item the complexity 
\end{itemize}
of either formulation. Regarding the update consistency:
\begin{rem}
Consider a current PWE configuration, $\left\{ f_{n}^{t}\in F_{n},\forall n\right\} $,
and a planned one, $\left\{ f_{n}^{t+1}\in F_{n},\forall n\right\} $,
set to be deployed at $t+1$. The update schedule, i.e., the order
of with which each tile $t$ will migrate from $f_{n}^{t}$ to $f_{n}^{t+1}$
requires consideration of transient effects.

For instance, consider that the first tile near a transmitter, $n_{1}$,
migrates to $f_{n_{1}}^{t+1}$ before all other tiles. The subsequent
propagation would be unintended, potentially yielding interference,
attenuation and eavesdropping potential. We denote this problem as
update consistency and study it in the ensuing section~\ref{subsec:Consistent-Update-Formulation}.
\end{rem}
Regarding the complexity:
\begin{rem}
A performance metric $\mu_{u}$ can be strongly dependent to a small
subset of $\left\{ f_{n}\right\} $ and independent to the rest.
\end{rem}
We consider the following supporting example. Let a PWE configuration
$\left\{ f_{n}\right\} $ leading to a metric $\mu_{u}$ for user
$u$, expressing his received power of the useful signal. The function
$\mu_{u}\left(\left\{ f_{n}\right\} \right)$ is strongly influenced
by only a subset of the total set of configured tiles which form a
cascading path by steering power from one to another. Changing any
of these functions completely (e.g., set to absorb, or steer somewhere
off-path) will result to an sharp reduction of $\mu_{u}$. On the
other hand, changing the EM Function of any tile in the PWE, which
does not receive impinging waves destined to $u$, will have no effect
on $\mu_{u}$.
\begin{rem}
A performance metric $\mu_{u}$ can be highly discontinuous with regard
to its strongly dependent tile functions. 
\end{rem}
The rationale follows the previous example, i.e., abrupt steering
changes at tile over a path lead to sharp decreases in the $\mu_{u}$
values, hence discontinuities in $\mu_{u}\left(\left\{ f_{n}\right\} \right)$.
However, altering a tile function to, e.g., yield a slightly different
efficiency degree $\epsilon$, as defined in relation~(\ref{eq:epsilon}),
leads to slight $\mu_{u}$ value changes and, hence, defines a relatively
small, continuous region of $\mu_{u}\left(\left\{ f_{n}\right\} \right)$.

These remarks support that solving the PWE configuration optimization
problem requires knowledge of the PWE graphs properties. (Here we
exclude simple cases that can be solved via exhaustive search, i.e.,
those with very few tile-, user-, and EM function-combinations). We
make note of some interesting such relations later, in Section~\ref{subsec:Graph-theoretic-challenges}
which outlines the related open challenges. 

\subsection{Consistent Update Formulation}\label{subsec:Consistent-Update-Formulation}

In this section, we formalize the problem from a consistent network
update perspective. The formalization of this section ensures that
path updates discussed in the next section (Section~\ref{subsec:Approximations-based-on})
are realized as they are envisioned, considering possible delays in
updates applied in each tile. In doing so, we provide a tutorial on
building a mixed integer linear program, which ensures consistent
network update, while minimizing the number of rounds, or the number
of updated tiles, as discussed in the previous section (Section~\ref{subsec:A-VNE-Embedding}).

We first point out that the problem at hand closely ties with the
similar problem that exists in the context of \emph{software defined
networks (SDNs)}. An SDN-based network design allows for a centralized
software program, to take the control of multiple data-plane elements,
allowing for a more flexible and efficient traffic engineering~\cite{ref50}.
However, such a centralized control comes with its own challenges,
in particular in terms of consistency of network updates: due to the
inherent differences between networking elements, they might perform
updates out of the intended order by the centralized control, which
might result into the formation of transient networking loops or black
holes~\cite{ref51,ref52}. As mentioned earlier, such
issues, in the context of PWE can result in multipath reverberation
or total signal loss. A suggested solution to keep updates consistent
is to perform updates in rounds, in such a way that any possible order
of updates in a round does not result in one of the aforementioned
problems~\cite{ref53}.

Another optimization objective that has been studied in the context
of SDNs, and closely ties with the optimization of parameter $R^{t}$,
is the number of switch interactions ($\texttt{touches}$)~\cite{ref54}.
Number of touches, in the context of PWEs, can be translated to the
number of tile interactions, which formed the basis of our formulation.

In the following formulation, we provide an exact Mixed Integer Linear
Program (MILP) formalization, to minimize the total number of interactions
with RIS units ($\texttt{touches}$), considering certain number of
rounds ($\texttt{rounds}$). This formulation utilizes the following
variables:
\begin{itemize}
\item For a tile $u$, we consider the binary variable $a_{v}^{t}$ to indicate
tile $u$ being active at round $t$. Additionally, for such a tile,
we consider an integer variable $o_{u}^{t}$, which indicates an order
for tile $u$ at round $t$.
\item For a link $\left(u,v\right)$ (there exists a link $\left(u,v\right)$
if tiles $u$ and $v$ have a LOS or nLOS), and round $t$, we define
an activity variable $a^{t}\left(u,v\right)$, which is one if both
$u$ and $v$ participated in a communication path at round $t$.
\item For a pair of tiles $u,v$, we define the integer variable $dis_{u,v}^{t}$
to denote distance between $u$ and $v$ at round $t$. Furthermore,
considering another tile $w$, we consider an auxiliary binary variable
$a_{u,v,w}$. 
\end{itemize}
For all the variables above, the $\left(s,d\right)$ in the superscript
of variables indicates that the variable is for transmitter $s$ and
receiver $d$. Furthermore, we assume that at the round $t$, we have
access to the list of transmitter-receiver pairs as $P^{t}$. In this
formulation, we consider $M$ to be a very large integer (at least
larger than the number of tiles).
\begin{algorithm}
\caption{Consistent Update Formulation}\label{Alg:=000020Update=000020Formulation}

\begin{algorithmic}[1]

\STATE \textbf{Minimize} $\texttt{touches}$ 

\STATE $\sum_{t=1}^{\text{\texttt{rounds}}}\sum_{u=1}^{n}c\left(a_{u}^{t},a_{u}^{t-1}\right)\le\texttt{touches}$
\label{line:=000020capped} 

\FORALL{$t<T$} 

\FORALL{$\left(u,v\right)\in E$} 

\STATE $a_{\left(u,v\right)}^{t}\le a_{u}^{t},a_{\left(u,v\right)}^{t}\le a_{v}^{t}$ 

\STATE $a_{\left(u,v\right)}^{t}\le dis_{u,v}^{t}$ 

\STATE $a_{\left(u,v\right)}^{t,\left(s,d\right)}\le a_{\left(u,v\right)}^{t}\hfill\forall\left(s,d\right){}^{t}\in P^{t}$ 

\ENDFOR 

\FORALL{$\forall u,v,w\in V$} 

\STATE $dis_{u,v}^{t}\le dis_{u,w}^{t}+dis_{w,v}^{t}$ 

\STATE $dis_{u,v}^{t}\ge dis_{u,w}^{t}+dis_{w,v}^{t}-M\cdot\left(1-a_{u,v,w}\right)$ 

\ENDFOR 

\FORALL{$\left(s,d\right){}^{t}\in P^{t}$} 

\STATE $\sum_{\forall\left(s,v\right)\in E}a_{\left(s,v\right)}^{t,\left(s,d\right)}=1$ 

\STATE $\sum_{\forall\left(u,d\right)\in E}a_{\left(u,d\right)}^{t,\left(s,d\right)}=1$ 

\STATE $\sum_{\forall\left(v,w\right)\in E}a_{\left(v,w\right)}^{t,\left(s,d\right)}=\sum_{\forall\left(w,u\right)\in E}a_{\left(w,u\right)}^{t,\left(s,d\right)},\,\forall w\in V\setminus\{s,d\}$ 

\ENDFOR 

\STATE $a_{\left(u,v\right)}^{t}+\frac{o_{u}-o_{v}}{|V|-1}\le1\hfill\forall\left(u,v\right)\in E$
\label{line:=000020MTZ}

\ENDFOR 

\end{algorithmic}
\end{algorithm}

The overall design of Algorithm~\ref{Alg:=000020Update=000020Formulation}
is as follows. It first ensures that the number of tile changes is
capped by $R$ . Here, we use the comparator $c$ defined in equation~(\ref{eq:comparator}).
We then ensure that the paths between a transmitter and receiver are
the shortest paths, and keep track of tiles that need to be active
in each path. In this formulation, we use a variant of Miller-Tucker-Zemlin
formulation~\cite{ref55} to ensure loop-freedom during the
update.

In order to speed up our MILP, in case a near-optimal result would
be sufficient, we can use a \emph{linear program relaxation}, and
then get an exact update order of tile updates through (randomized)
rounding.

\subsection{Approximations based on Heuristics and Path finding Algorithms}\label{subsec:Approximations-based-on}

In the preceding Sections, it was shown that a formal PWE optimization
can be computationally intensive. Therefore, approximate but efficient
heuristics has been proposed in the literature and are surveyed below. 

\subsubsection{Path Finding-driven Approximations}\label{subsec:Path-Finding-driven-Approximatio}

Consider a set of users in a PWE graph. Let the tiles in the LOS vicinity
of each user be denoted as \emph{first-contact} tiles. The path-finding
approach involves finding routes that connect the first-contact tiles
of user~$1$, to the first contact tiles of user~$2$. This can
also be generalized for multiple users. 

A straightforward approach is to model this objective as a k-shortest
paths problem~\cite{ref56}, using the link loss (derived
by the link distance and any consequences of Remark~\ref{rem:There-exist-visibility})
as the cost metric. The intuition behind this method is that by utilizing
multiple paths, the received signal power can be accumulated, while
minimizing delay helps reduce delay spread and power loss, especially
in environments with abundant path diversity. This approach can be
extended by incorporating a more sophisticated cost function that
captures additional performance factors. 

Although traditional k-shortest or similar path algorithms can be
applied, they must be adapted to enforce certain constraints that
may be set. Major factors include:
\begin{itemize}
\item Using only up to a specific number of EM functions per metasurface
simultaneously, which puts a limit to the number of air routes that
can intersect at any given graph vertex.
\item A user set prioritization policy, such as finding paths for the most
distant user pairs first~\cite{ref8}, since initial
path finding attempts are likely to yield better results, compared
to when parts of the graph have been already marked as occupied by
EM functions.
\item Filtering the possible paths further, e.g., by avoiding links near
possible eavesdroppers, or by requesting that the final link be perpendicular
to the mobility trajectory of the receiver, in order to mitigate Doppler
effects in mmWave communications. (An example is evaluated later,
in Section~ \ref{sec:Evaluation:-A-Factory}).
\end{itemize}
Furthermore, while the k-shortest paths formulation offers a reasonable
starting point, more advanced formulations are needed to fully capture
the multi-objective nature of the problem. For instance, one can adopt
a multi-objective optimization framework that treats delay and signal
strength as separate objectives. This allows for a more nuanced exploration
of the trade-off between these often-conflicting goals. Achieving
this requires more sophisticated algorithms capable of handling multiple
objectives such as multi-objective heuristic optimizers, Pareto optimization
methods, or scalarization techniques like the weighted sum approach.
The latter can reduce the multi-objective problem to a single-objective
shortest path problem by combining objectives into a single cost function.

We also note that the optimization of the number of updated tiles
has close ties with the problem of \emph{lexicographical shortest
path}~\cite{ref57}, in which the shortest path
should be determined based on multiple objectives. The algorithms
for this problem are an extension of shortest path algorithms that
use the \emph{best-first search} approach~\cite{ref58} (similar
to the Dijkstra algorithm). In the context of our problem, we briefly
describe a greedy algorithm that works as follows: Initially, it sorts
transmitter-receiver pairs based on their preserved distances, in
an ascending manner. It also initializes a set to keep track of tiles
that have been part of the shortest path. Then, for each transmitter-receiver
pair, we build the shortest path as follows: we run a Dijkstra algorithm,
with the difference that in each step, if two tiles have the same
distance, we select the one that has already been part of the shortest
path before.

\subsubsection{Propagating Heuristic-driven Approximations}

Propagating heuristics are a vast class of approximate solvers, which
rely on the progressive exploration of a solution space~\cite{ref59}.
They can be categorized as back-propagating or front-propagating,
depending on their direction of operation. Front-propagating initiate
from a known initial state and attempt to reach an optimal one, while
back-propagating heuristics start from the required, optimal conditions
and attempt to find initial states that correspond to them. In the
following, we provide a brief tutorial on their application to the
problem of PWE optimization. 

\paragraph{Back-Propagating}

\begin{figure}[!t]
\centering{}\includegraphics[width=1\columnwidth]{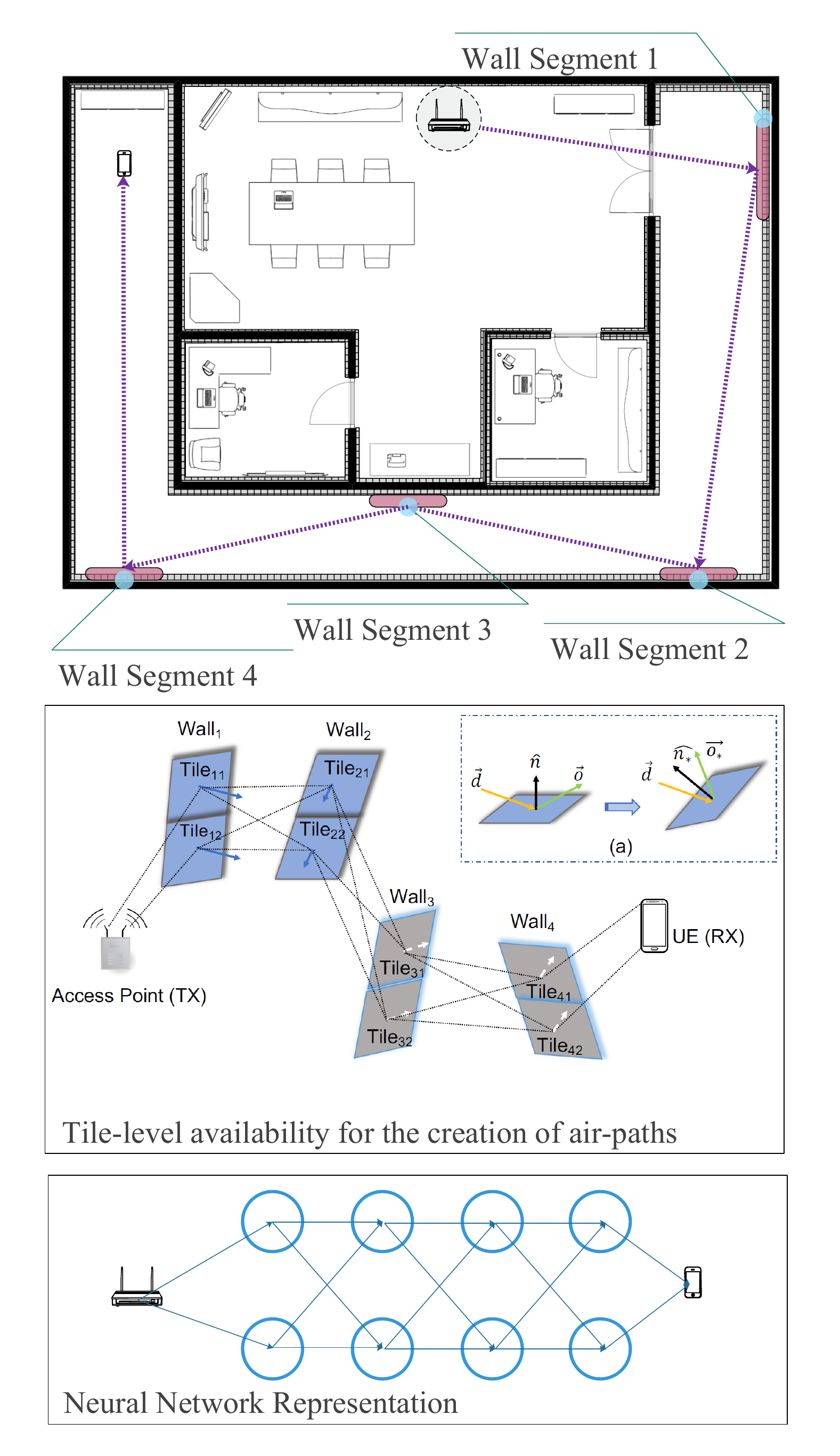}\caption{Overview of a back-propagating heuristic for PWE optimization. Top
inset: an exemplary floorplan where the walls are covered with metasurface
tiles. Middle inset: A selection of wall segments is made, which will
connect two wireless devices via wave propagation engineering. Bottom
inset: a corresponding neural network is constructed, which seeks
to attained a required level of performance at the receiving end.
Inset (a): once the back-propagation process is complete, each neuron
is mapped to an EM Function depending on the similarity between the
neuron's weights and the EM scattering profiles that can be attained
by the corresponding metasurface.}\label{fig:nnpaths}
\end{figure}
This direction is expressed via the concept of using neural networks
to optimize wireless communication systems by making the architecture
directly interpretable~\cite{ref60}. Each neuron in the
network corresponds to a specific part of the system being modeled,
providing a direct correspondence with how the network processes information.
The state of a neuron directly reflects the EM Function of a metasurface
tile, allowing for an automated mapping of the state of a trained
neuron to an EM Function that should be enabled. The concept is visualized
in Fig.~\ref{fig:nnpaths}.

The number of layers and nodes in each layer is determined by the
number of walls with metasurface tile coating that a transmitted wave
will impinge upon before reaching its intended receiver. In other
words, a rough per-wall routing decision is made first, using classic
algorithms such as Dijkstra or A{*}~\cite{ref59,ref61}.
(Notice that this employs a different graph than the one described
in section~\ref{subsec:A-Graph-Model}). This approach ensures that
the network captures the general layout of the physical environment.
The input layer represents the initial conditions at the transmitter
side, while the output layer represents the required, optimal conditions
expected at the receiver side.

A multi-layer perceptron neural network is employed because it enforces
similar activation and actuation of its neurons, aligning with how
tiles manipulate the flow of electromagnetic waves from a transmitter
to a receiver. Back-propagation is used to adjust the nodes based
on the received power pattern at the output layer in the previous
feedforward step. A linear ramp is used as the node activation function
to align with energy conservation principles of wireless propagation.

\paragraph{Front-propagating heuristics}

\begin{figure}
\begin{centering}
\includegraphics[width=1\columnwidth]{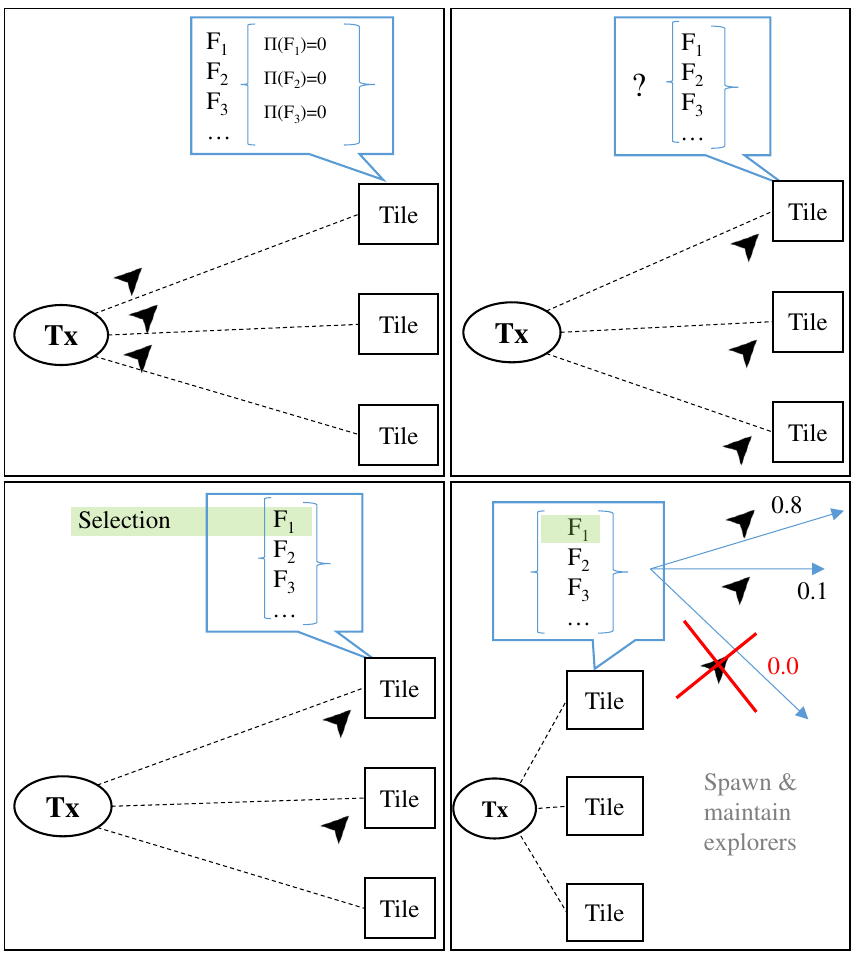}
\par\end{centering}
\centering{}\caption{An example of a forward-propagating heuristic for configuring PWEs.
Top left: initialization. Top right and bottom left: stochastic EM
selection based on the explorers' accumulated knowledge. Bottom right:
Spawning and maintaining explorers based on the EM selection at a
metasurface tile.}\label{fig:ants}
\end{figure}
Forward propagating heuristics assume a set of solution space-exploring
processes, which can either be centrally coordinated or distributed
in their workflow. Initially, each searcher explores in random directions.
When one searcher discovers a promising solution, it retraces its
path back to the starting point, leaving behind heuristic information
to guide future searches. This information influences other searchers
to follow the same path with greater likelihood, leading to convergence
on an optimal solution. This process is a form of reinforcement learning
and serves as a flexible optimization framework with various applications.

In the context of PWE optimization, this established optimization
process cannot be directly applied, as the optimization parameters
are not the links and weights to be selected. Instead, the challenge
lies in selecting the appropriate EM Function for each metasurface.
This selection alters the attributes of multiple links connected to
the corresponding metasurface, depending on the direction of the impinging
wave. This results in a dynamic optimization problem in which the
graph changes as a wave propagates through it.

To address this novel graph routing problem, a forward-propagating
heuristic process can exemplary operate as illustrated in Fig.~\ref{fig:ants}
\cite{ref62}. The process begins with explorers being
released from the transmitter and traveling along each link towards
the nearest metasurface (top left panel). When an explorer reaches
a free metasurface, it selects one of the available EM functions randomly,
using their associated heuristic information as a probability distribution
function, as shown at panels at row $1$, column $2$ and row $2$,
column $1$. (This does not imply an exhaustive listing of all possible
EM functions, but just the selection of one possible EM Function).
Initially, this distribution is uniform but is updated throughout
the process. Once an explorer activates a function, it is destroyed
but also generates a group of new explorers based on the corresponding
output distribution, in order to capture the dissipating nature of
EM waves. Each new explorer carries information about the cumulative
propagation delay experienced so far and the remaining wave power.
Notably, any explorers whose remaining wave power falls below a predetermined
threshold are discarded, preventing unnecessary processing time. The
process then repeats for all other explorers in the graph. 

It is important to note that explorer advancement can be performed
centrally in rounds (i.e., all explorers are advanced by $1$ hop
per round until the process concludes), or completely autonomically.
Once an explorer reaches a receiver, the explored wave route is kept
in a list maintaining the top-$n$ paths in terms of power. When all
explorer have concluded, a wave route selection takes place, in order
to optimize the value of the selected metric. The process can accommodate
multiple receivers and transmitters. In such cases, explorers carry
an identifier denoting their intended receiver to distinguish between
useful signals and interference. 

\section{The PWE integration to Communication Systems}\label{sec:The-PWE-integration}

We proceed to describe the control workflow of PWEs, and their integration
to existing networks and resource allocation systems.

\subsection{The PWE Workflow as a Networked System}

\begin{figure*}[t]
\centering{}\includegraphics[width=0.8\paperwidth,viewport=0bp 0bp 796bp 635bp]{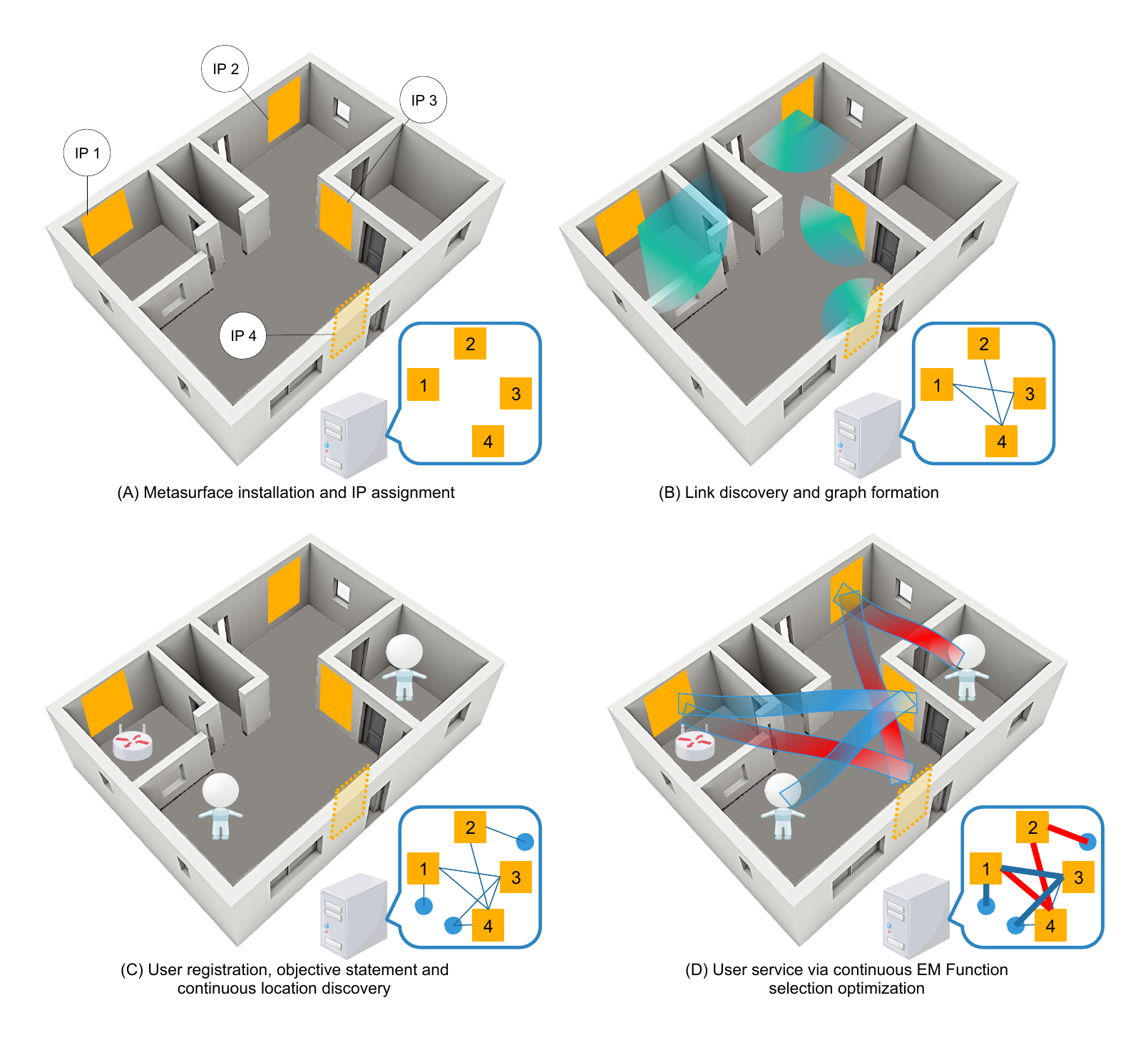}
\caption{ \textcolor{black}{The four major steps of the PWE workflow: a) Metasurface
units are physically installed and receive Internet Protocol (IP)
addresses. b) The metasurface linkage is detected following a manual
entry or automated process (beam scanning). c) Users register to a
web service and state their objectives within PWE. d) The user locations
are continuously tracked and the PWE is orchestrated accordingly,
following a graph optimization algorithm.}}\label{fig:PWEworkflow}
\end{figure*}

The operating workflow of a PWE involves a dynamic interplay between
user devices, metasurfaces, the existing network infrastructure and
a central orchestration service~\cite{ref63}. Figure~\ref{fig:PWEworkflow}
outlines four major steps comprising this workflow as follows:

First, the metasurfaces are physically installed on walls, ceilings,
etc., and connect to the existing local network, receiving IP addresses
and advertising their availability and operating status to an orchestrating
server, as shown in Fig.~\ref{fig:PWEworkflow}-a. (Notice that the
orchestrating server is a software process, which can be hosted within
existing computers, routers, or access points). 

Second, and at least once after their installation, the metasurfaces
are represented within the orchestrating process in the form of the
aforementioned graph. This process is semi-automatic. Akin to common
smart devices, the metasurfaces are registered to parts of a floorplan
in a coarse manner. An extensive beam scanning and steering process
follows (Fig.~\ref{fig:PWEworkflow}-b), with the objective of detecting
line-of-sight visibility (i.e., link existence) and link characteristics
(i.e., attenuation and delay) among metasurface pairs. Metasurfaces
can be equipped with assistive circuitry, such as infra-red transceivers
to efficiently perform this task~\cite{ref64}.
The objective of this process is not just to derive the coordinates
of the locations of metasurfaces, but rather to create a graph representation
of the PWE within the orchestrator's software process.

Third, users are registered to the PWE and express their performance
objectives. Users become aware of nearby PWEs through broadcast messages
advertising its existence. These announcements facilitate device-to-environment
association using shared authorization mechanisms. This interaction
enables PWEs to recognize and accommodate the presence of user devices
within their vicinity. Upon this initialization process, users state
the service objectives they wish to pursue (e.g., enhanced security,
wireless power transfer, quality of service).

Notably, the user initialization also involves his continuous tracking
by an existing user localization system operating independently within
the PWE. Exemplary approaches include optical tracking~\cite{ref65},
or RF-driven tracking employing artificial intelligence~\cite{ref66}.
Notably, metasurfaces exhibit inherent capabilities that enable them
to function as user localization devices themselves~\cite{ref67,ref68,ref69,ref70}.

Users in service maintain an open control channel to the orchestrator
to relay updated objectives and performance metrics, and receive beamfor\textgreek{μ}ing
instructions for their devices, in order to optimally match their
RF emissions to the status and location of the metasurfaces. (To manage
these transitions, mechanisms adapted from Session Initiation Protocol
can be employed~\cite{ref71}, streamlining communication
between devices and PWEs).

Finally, the orchestrator executes a graph optimization algorithm,
as described in the previous section, and deduces the optimal EM Functions
for each metasurface, matching the user objectives and positions.
These EM Functions are relayed to the metasurfaces via the IP network,
hence becoming \emph{deployed}. The PWE enters a continuous cycle
of receiving new user registrations, new user objectives or updated
user positions and maintaining the EM Functions deployed to each metasurface. 

\begin{figure}[!t]
\begin{centering}
\textsf{\includegraphics[clip,width=1\columnwidth]{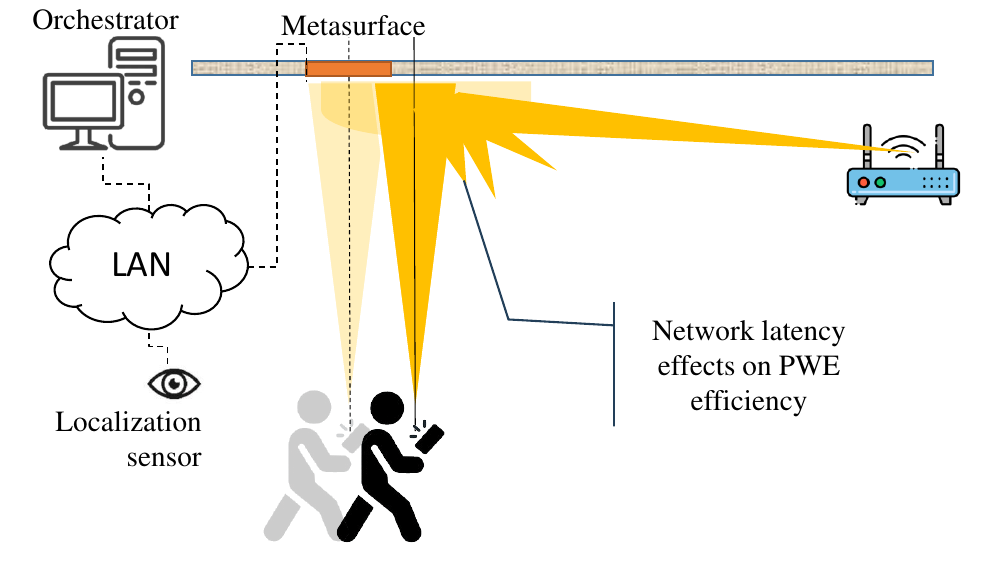}}
\par\end{centering}
\caption{Overview of the PWE adaptation workflow, where a server that continuously
monitors wave emissions from users and dynamically configures the
metasurfaces. The performance of the network connecting the control
elements of the PWE has important effects on its efficiency.}\label{fig:setupMobility}
\end{figure}
To optimize PWE configuration effectively, the orchestrator needs
to accurately detect the wavefronts emitted from user devices. This
can be achieved through various methods, including the use of built-in
sensors within metasurfaces or by leveraging compressed sensing principles
in a sensor-less approach. 

However, in mobility scenarios, as illustrated in Fig.~\ref{fig:setupMobility},
delays in completing the sense/process/configure cycle can lead to
misaligned user emissions and metasurface configurations, ultimately
resulting in sub-optimal wave routing~\cite{ref72}.
Thus, the network infrastructure and protocols involved in the communication
between the orchestrator, the user localization systems and the metasurfaces
is of critical importance, introducing interesting research challenges
discussed later, in section~\ref{sec:Discussion-and-Network}.

\subsection{Interfacing with Existing Networks: A Standardization Driven Aspect}\label{subsec:Katsalis}

The standardization landscape for metasurfaces, mainly via their RIS
variation, is rapidly evolving and involves multiple complementary
organizations and forums addressing different aspects of the technology,
in order to interface it with the existing network infrastructure.
Three major integration directions are presently studied, each via
a corresponding set of bodies:
\begin{enumerate}
\item ETSI ISG (Industry Specification Group) RIS leads the effort by defining
metasurface fundamentals, architectures, channel models, testing methodologies,
and management aspects to prepare the ground for future normative
standards. 
\item ETSI ISC (Integrated Sensing and Communication) RIS consider metasurfaces
within broader studies on communications evolution towards 6G, capitalizing
in Integrated Sensing and Communication (ISAC)~\cite{ref19}.
IEEE contributes through research-driven standards and publications,
particularly in antennas, propagation, and measurement techniques.
(In a similar but independent manner, ITU-R and 3GPP provide spectrum-
and propagation-related frameworks relevant to metasurface deployment
and interfacing with specific systems such as Non-Terrestrial Networks
and Radio Access Networks~\cite{ref73}).
\item ETSI ISG NFV (Network Functions Virtualisation) investigates the use
of metasurfaces as a cloud resource. 
\end{enumerate}
Together, these sets of bodies form a coordinated standards ecosystem
shaping the industrialization and future integration of metasurfaces
into next-generation wireless networks. We proceed to survey the technical
integration focus and activities of each.

\textbf{ETSI ISG RIS} is an Industry Specification Group within ETSI
focused on enabling programmable and controllable radio environments
through metasurfaces. The group develops use cases, architectures,
channel models, testing frameworks, and management concepts for RIS,
targeting integration into 5G-Advanced and future 6G networks. Over
a series of documents~\cite[ETSI GR RIS 001-008]{ref74}, this group
has surveyed the following topics:
\begin{itemize}
\item Architectures, operating modes, and frequency ranges for future wireless
networksalong with key-use cases where metasurfaces can enhance coverage,
spectral efficiency, beam management, physical-layer security, localization,
sensing, and energy efficiency. Additionally, deployment scenarios
across indoor, outdoor, hybrid, static, and nomadic environments are
considered. Control plane options and outlines technical, interoperability,
and regulatory requirements are discussed. 
\item Metasurface-internal architectures, control modes, hardware structures,
and fabrication options, highlighting trade-offs between flexibility,
complexity, and performance. 
\item Communication and channel models for metasurfaces, covering both near-field
and far-field regimes. It defines deterministic and statistical channel
models, including path-loss, multipath, interference, reradiation,
and polarization effects across sub-6 GHz and high-frequency bands.
Metasurface-aided device localization and sensing models, including
SISO, MIMO, metasurface-aided, and metasurface-standalone scenarios.
Channel estimation methodologies and key performance indicators (KPIs)
for quantifying the metasurface capabilities and performance are also
discussed, along with robust performance specifications. 
\item Explicit metasurfaces (as opposed to simple RIS) capabilities are
briefly discussed, including transmission, sensing, analog computing,
and caching.
\end{itemize}
In other words, the first integration consideration, studied by ISG
RIS, is to provide a clear specification of the physical layer of
metasurfaces, their associated performance metrics and signal processing
techniques relating to wireless channel models. 

\textbf{ETSI ISG ISC} is an Industry Specification Group within ETSI
that focuses on the joint design and integration of communication
and sensing capabilities. As metasurfaces can shape, reflect, refract,
or focus electromagnetic waves, enabling the network to create favorable
propagation conditions for both data transmission and sensing, ISAC
systems extend their sensing coverage, enable virtual line-of-sight
links, and improve detection, localization, and tracking accuracy,
especially in NLOS or obstructed environments. Metasurfaces act as
passive or semi-passive sensing enablers, creating additional sensing
paths and angles without adding active radio transmitters. From an
architectural perspective, metasurfaces can be coordinated with ISAC
transmitters and receivers to jointly optimize beamforming, waveform
design, and resource allocation for both functions.

In this context, the ISC group develops use cases, requirements, architectures,
and enabling technologies for ISAC, targeting applications such as
localization, environment monitoring, healthcare, transportation,
and smart infrastructure~\cite{ref75}. ISC considers metasurfaces
as a tightly integrated technology that enhances and extends the capabilities
of existing communications infrastructure, rather than as a standalone
sensing system. 

Overall, the ISC approach is more macroscopic compared to ISG, focusing
on higher-level but modular functions and single tasks that can be
performed by a metasurface to the benefit of either a user or another
network sub-system.

Finally, the \textbf{ETSI ISG NFV} focuses at an even higher level,
and defines the architectural framework, interfaces, and management
principles for virtualizing network functions on standardized cloud
infrastructure. Network Function Virtualization NFV is an architecture
concept in telecommunications and networking where network functions---that
traditionally ran on dedicated, proprietary hardware---are instead
implemented as software running on standard (commodity) servers~\cite{ref76}.
ETSI ISG NFV has investigated the case of managing RIS as an additional
cloud computing resource.

In this aspect, metasurfaces are considered as a new type of NFV Infrastructure
(NFVI) resource that extends the traditional compute--storage--network
model toward programmable physical radio environments~\cite[ETSI GR NFV-EVE 023]{ref76}.
Specifically, ISG NFV treats metasurfaces as a physical network resource
that can be onboarded, inventoried, and managed within the NFV framework,
similarly to other infrastructure components, via the Physical Infrastructure
Manager (PIM). Metasurfaces are seen as capable of enhancing coverage,
signal quality, and network efficiency by actively controlling radio-wave
propagation, effectively extending the radio access closer to end
users. Two alternative integration models are defined. In the first
model, metasurfaces are managed as infrastructure elements abstracted
and exposed to the management system for orchestration and resource
allocation. In the second model, metasurfaces are modeled as Physical
Network Functions where a dedicated descriptor defines their connectivity
and role. This body also highlights key management challenges, including
the metasurface lifecycle management, monitoring (e.g., energy consumption),
interaction with external controllers (often owned by third parties
such as building owners), and defining clear demarcation points between
the cloud management system and external network management systems. 

\section{Simulating Networked PWEs: The Omnet++ case}\label{sec:Simulating-Networked-PWEs:}

Simulators supporting metasurfaces at the physical layer have emerged,
both commercial and open-source~\cite{ref77,ref78}.
Moreover, RIS-specific simulators have also appeared~\cite{ref79}.
These tools operate on either ray-tracing of full-wave solver principles
and provide signal quality indications, assuming that the metasurfaces
in the system have been configured (i.e., after the optimization problem
has been solved) and the user device parameters (such as directivity,
power, sensitivity) have been set. 

Conceptually, one could employ these tools within a heuristic optimization
loop, executing simulations and improving upon the choice of EM Functions
per tile. However this entails a computational burden that may not
be generally sustainable, making these tools more fit for the final
step of an evaluation, deducing more accurate results for an already
optimized PWE configuration decision. (This also affects optimal metasurface
deployment studies~\cite{ref63}, which require an
even larger simulation overhead). 

It is for the same reason that these tools do not merge directly with
network-layer concerns and simulations. However, network-layer simulators,
such as NS, Omnet++ and OPNET are known to effectively cover wireless
standards, employing more computationally scalable approaches, such
link budget tools and statistical communication models~\cite{ref80}.
Such approaches have already started to appear and gain momentum in
the area of vehicular networks~\cite{ref46}. In the
following, we outline the process of incorporating metasurfaces into
any discrete event simulator for networks, using Omnet++ as a driving
example~\cite{ref81}.

\textbf{Prerequisites.} The simulation design will rely on the PWE
graph model. In this model, the EM Functions are well-defined, using
the graph links as EM Function parameters as required. In general,
a tile can treat any subset of its links as incoming wave directions,
and any subset as departing wave directions. We define the following
problems:
\begin{problem}
How does a graph-level EM Function translate to an actual EM performance?
In other words, following the notation of Fig.~\ref{fig:graph},
how does one obtain the efficiency degrees per departing wave direction?
\end{problem}
\begin{sol}
Decompose each complex EM Function into simple ones, for which a codebook
exists or that can be calculated in reasonable time. Each complex
EM Function can be effectively decompose to multiple single steerings,
if from one incoming link to one departing link. If a codebook exists
for these single steerings, then the efficiency degrees for the complex
function can be derived following the process described in the context
of Fig.~\ref{fig:merge}. Deriving the codebook for single steerings
can be accomplished via existing standalone tools~\cite{ref82,ref83}.
\end{sol}
\begin{problem}
What properties should a PWE graph have?
\end{problem}
\begin{sol}
In general, several metasurfaces tiles are placed over walls and ceilings.
These tiles are node graphs without connectivity. However, walls and
ceilings composing a room comprise tile sets with extensive connectivity,
as shown in Fig.~\ref{fig:graphs}. Further discussion on the topic
can be found in Section~\ref{subsec:Graph-theoretic-challenges},
while graph generation and visualization scripts can be found online~\cite{ref84}.
\end{sol}
\begin{problem}
How does a wave attenuate over a PWE graph?
\end{problem}
\begin{sol}
Consider a path originating from a transmitter and ending at a receiver
within a PWE graph, crossing $N$ tiles and $N+1$ links, effectively
over $N$ steering EM Functions. If the tiles can employ the collimation
functionality (cf. Section~\ref{subsec:collimate}) in tandem with
steering, then the received power will be (extending ~\cite{ref46}):
\begin{equation}
P_{r}=\frac{P_{t}\epsilon_{t}\epsilon_{r}\prod_{n=1}^{N}\epsilon_{n}}{\prod_{n=1}^{N+1}\left(\frac{4\pi f}{c}\right)^{2}\left\Vert l_{n}\right\Vert ^{\alpha_{n}}}\cdot\prod_{n=1}^{N+1}L_{n},\label{eq:productpl}
\end{equation}
where $P_{t}$ is the transmission power, $\epsilon_{t}$ and $\epsilon_{r}$
the antenna efficiency factors of the user devices along the directions
of the first and the last link, $\epsilon_{n}$ is the EM Function
efficiency at tile $n$, $\left\Vert l_{n}\right\Vert $ is the distance
of the $n^{th}$ link over the path, $a_{n}$ is a factor expressing
the path loss as affected by the efficiency of the collimation functionality
and value of $\left\Vert l_{n}\right\Vert $, and $L_{n}$ is any
link loss stemming from partial LOS as stated in Remark~\ref{rem:There-exist-visibility}
or~$1$ otherwise. (While $a_{n}$ can be statistically derived by
the physical-layer simulation tools, its value should be less than~$2$,
which is the free space loss exponent. A rule of at thumb would be
to assume $a\approx1$ near a tile and $a\approx2$ in its far field). 
\end{sol}
When the tiles do not support collimation functionalities, then they
essentially act as plain reflectors (albeit without mechanical movement).
In this case, equation (\ref{eq:productpl}) is rewritten as~\cite{ref46}:
\begin{equation}
P_{r}=\frac{P_{t}\epsilon_{t}\epsilon_{r}\prod_{n=1}^{N}\epsilon_{n}}{\left(\frac{4\pi f}{c}\right)^{2}\left(\sum_{i=1}^{N+1}\left\Vert l_{n}\right\Vert \right)^{\alpha}}\cdot\prod_{n=1}^{N+1}L_{n},\label{eq:sumpl}
\end{equation}
where $a$ is now uniform and takes a value of less than $2$, as
dictated by the type of the 3D setting~\cite{ref85}. 

Notably equations (\ref{eq:productpl}) and (\ref{eq:sumpl}) can
be combined accordingly when using mixed-technology metasurfaces within
the same PWE. 
\begin{rem}
Having a defined PWE graph and EM Functions (e.g., via the processes
described in Section~\ref{subsec:Approximations-based-on}), as well
as the user device parameters, one can traverse the PWE graph and
derive the resulting approximate Power Delay Profile.
\end{rem}
\textbf{Omnet++ integration.} As discussed in Section~\ref{sec:PWE-Performance-Optimization},
the PDP is a set containing the power, latency, phase (and polarization)
of each path reaching the receiver. It is noted that dedicated EM
Functions can modify the phase and polarization of a wave, essentially
attaining over the air path equalization and polarization matching,
respectively. In any case, the PDP contains all the necessary information
to derive the resulting bit error error rate and fading model, given
a selected modulation scheme and a symbol duration and, hence, the
attainable data transfer rate as well~\cite{ref86}. 

Therefore, PDP will act as the intermediate to interface the PWE world
to the Omnet++ simulation engine. 

Omnet++ comes tightly integrated with the INET/MANET framework, which
contains an extensive set of protocols per OSI layer stack, including
wireless propagation. The latter can be approached either: 
\begin{itemize}
\item from the more abstract level of link budget and stochastic channel
models (employing the classes $\texttt{WirelessChannel}$, $\texttt{IRadio}$,
$\texttt{RadioMedium}$, $\texttt{AnalogModel}$, $\texttt{Signal}$,
$\texttt{PacketLevelRadio}$ vs. $\texttt{AnalogRadio}$, $\texttt{PropagationLossModel}$,
$\texttt{PathLoss}$, $\texttt{Fading}$).
\item from the more precise waveform-level of simulation, providing ways
to simulate specific waveform encoding, decoding, modulation and distortion
(employing classes such as $\texttt{TransmissionBase}$, $\texttt{IRadioSignal}$
and $\texttt{AnalogModelResult}$). 
\end{itemize}
Since we have motivated the need for computationally lighter simulations,
the first approach is deemed as more fitting. In order to proceed
with it an new type of channel needs to be expressed, which will capture
the prerequisites listed above. This can be achieved by extending
the $\texttt{RadioMedium}$, which will allow for: 
\begin{itemize}
\item intercepting Omnet++ transmission events, 
\item transform the signal as dictated by the PWE graph, and finally
\item deliver it to the receiver, creating a modified copies of the original
Transmission/Signal object, in order to create the proper PDP.
\end{itemize}
Note that there are two approaches for modeling the PWE environment
within Omnet++:
\begin{enumerate}
\item Keep only the user devices present within the INET network, while
the actual tiles reside outside Omnet engine, providing a PDP that
matches the EM Function choices upon callback.
\item Model both the users and the tiles within the INET network.
\end{enumerate}
The first approach keeps most of the INET modeling approach intact
and requires only the introduction of a rather minor class extension.
Moreover, the existing tools for PWE graph generation, configuration
and traversal can be exploiting directly via simple interfaces. Additionally,
assuming that these Omnet-external tools can be called upon as services,
a high degree of parallelism can be achieved, with Omnet++ calling
upon multiple PWE graph configuration to evaluate in parallel, e.g.,
in terms of TCP-over-PWE performance. 

The second approach calls for re-implementing existing tools within
the Omnet++ codebase. It can offer a more mature approach in the long-term,
as the ensuing modularity would allow for targeted code improvements
and extensions, all exposed to the end user via the NED scripting
language, which is employed by Omnet++ in order to create scenarios
over an existing, compiled version of the INET/MANET framework.
\begin{figure*}[!t]
\begin{centering}
\textsf{\includegraphics[viewport=0bp 0bp 1073bp 790bp,clip,width=1\textwidth]{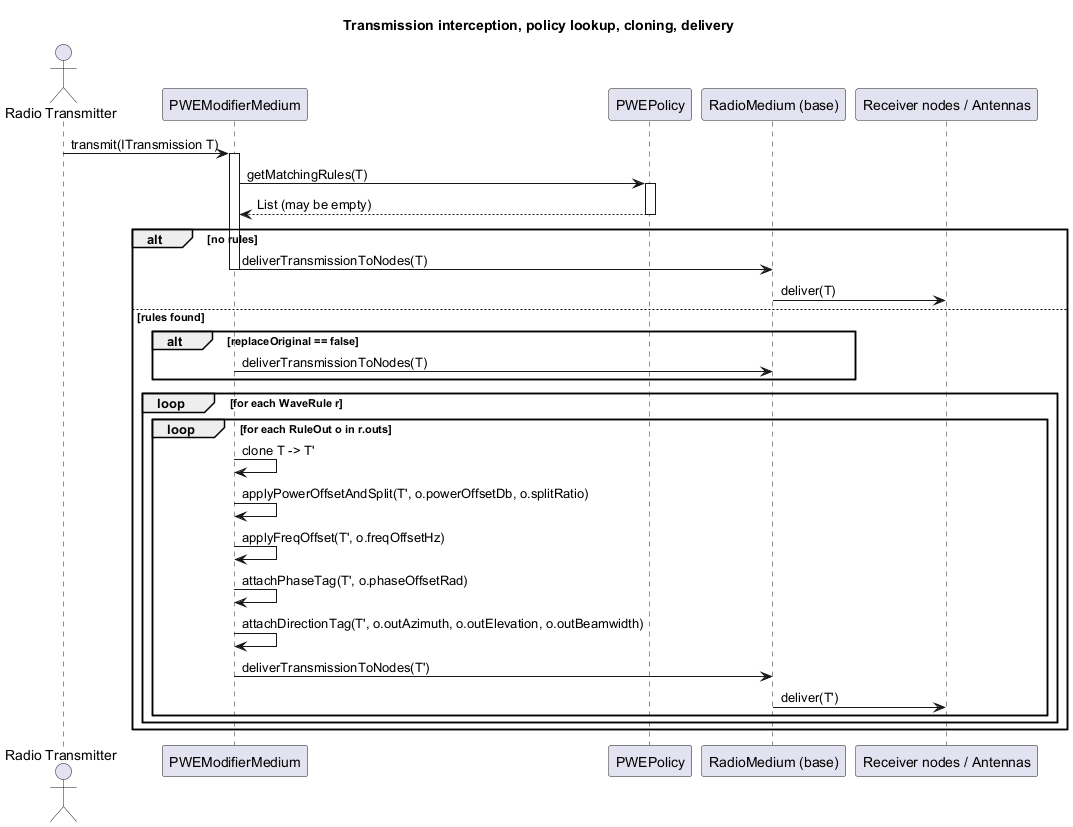}}
\par\end{centering}
\caption{UML overview of the proposed PWE-Omnet++ integration.}\label{fig:UML1}
\end{figure*}

An overview of the proposed implementation in UML format is given
in Fig.~\ref{fig:UML1}. We consider the class $\texttt{PWEModifierMedium}$
which extends $\texttt{inet::physicallayer::RadioMedium}$, intercepting
the INET reception/delivery hooks. (E.g., $\texttt{deliverPacketToChannel}$
/ $\texttt{deliverTransmissions}$ or $\texttt{interceptTransmission}$
/ $\texttt{handleTransmission}$).

Additionally, we introduce a custom singleton class called $\texttt{PWEPolicy}$,
which is a data holder for the communication objectives per user in
the system, as described in Section~\ref{subsec:A-VNE-Embedding}.
When a user emits a signal, it is intercepted by the $\texttt{PWEModifierMedium}$
and a lookup for corresponding objectives is made in the $\texttt{PWEPolicy}$
instance. If no entry is found, the wireless propagation occurs with
the PWE as-is (i.e., configures, partially configured or unconfigured,
i.e., regular propagation). 

If matching rules are found, then the simulation can either: greedily
reconfigure the PWE via the external service, leaving the rest of
the PWE configuration as-is, as a more lightweight simulation solution,
or proceed to completely reconfigure the whole PWE for all users and
user objectives. In any case, a PDP is produced which is passed to
the Omnet++ engine and dictates the number and parameters of all copies
of the original transmissions to be delivered to the Omnet++ receivers.
All other OSI layers simulated by Omnet++ operate normally. 

It is noted that, while many of the above specifics refer to Omnet++,
the same logic--i.e., calling upon PWE graph processing utilities
as external services--can be employed for other network simulators
(e.g., NS or OPNET) by using a similar interception at the class representing
the wireless propagation medium. 

\section{Evaluation: A Factory Application Example}\label{sec:Evaluation:-A-Factory}

\begin{figure}[h]
\centering{}\includegraphics[width=1\columnwidth]{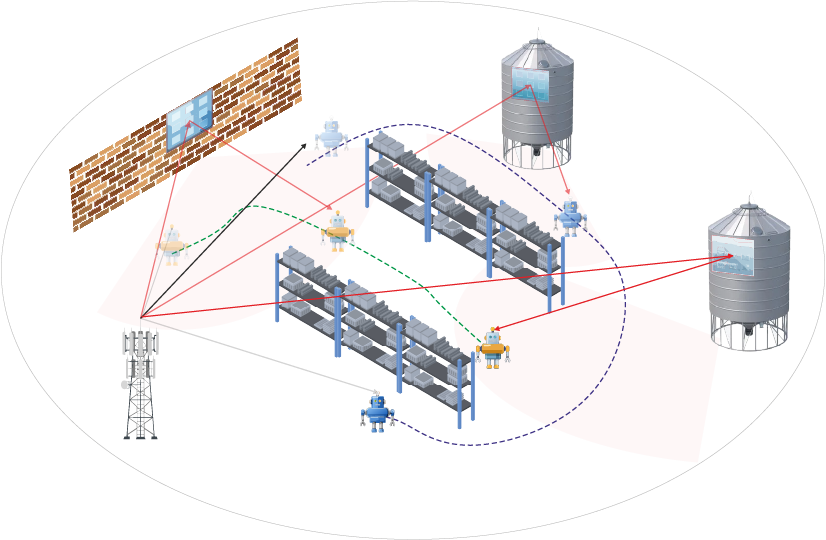}\caption{Conceptual application of a PWE in a factory setting.}\label{fig:factory}
\end{figure}

One of the main use cases envisioned for PWEs are in the context of
factory settings, exemplary shown in Fig.~\ref{fig:factory}. These
scenarios often have complicated wireless environments due to heavy
production machinery or storage racks blocking signals. Instead of
deploying multiple repeaters or access points, PWEs can be used to
extend coverage to areas that would otherwise be out of the line of
sight of the base station and, most importantly, mitigate propagation
phenomena that could be detrimental to strict machinery requirements,
such Computerized Numerical Control (CNC) equipment~\cite{ref87}.
Such systems require communication time-sensitivity and near-zero
error tolerance in their actuator-sensor control loops, especially
when operating in orchestration with other equipment. Additionally,
as the user devices in such environments often move on predetermined
and scheduled paths, such as with robots taking items from one rack
to another, it simplifies the measures required to control a PWE in
real-time.

In a more concrete scenario, robots move on pre-scheduled routes to
retrieve items from storage racks in a warehouse. They take these
items and move them to an area where they will be packed for shipping.
While the robots are moving, they still need to stay connected to
the network the entire time to transmit telemetry data as well as
receive any possible updates to their routing etc. PWEs can be deployed
to ensure that such constraints, and their networking requirements,
are met. Furthermore, as the robots are moving at high speeds, the
metasurfaces can be used to reduce the Doppler effect that would otherwise
occur, by ensuring that a wave meets the trajectory of a robot perpendicularly.

We proceed to study this setting in the following simulation, employing
the path finding-driven heuristic outlined in Section~\ref{subsec:Path-Finding-driven-Approximatio}.
Specifically, we employ the \noun{KpConfig} heuristic and the associated
simulation engine~\cite{ref8}. Physical-layer parameters
are summarized in Table~\ref{tab:TSimParams}.
\begin{table}[t]
\centering{}\caption{\foreignlanguage{english}{\textsc{Simulation parameters}}}\label{tab:TSimParams}
\begin{tabular}{|c|c|}
\hline 
Ceiling Height & $3$~$m$\tabularnewline
\hline 
SDM Dimensions & \textbf{$1\times1\,m$}\tabularnewline
\hline 
SDM Design & \cite{ref78}\tabularnewline
\hline 
Frequency & $60\,GHz$\tabularnewline
\hline 
Tx Power & $30\,dBm$\tabularnewline
\hline 
\multirow{2}{*}{Tx/Rx Antenna types} & Pyramidal horn, $80^{o}$ beamwidth\tabularnewline
 & (pointing upwards)\tabularnewline
\hline 
Max ray bounces & $50$\tabularnewline
\hline 
Min considered ray power & -$250$~dBm\tabularnewline
\hline 
\end{tabular}
\end{table}

\subsection{Simulation Scenario and Results}

\begin{figure}[t]
\begin{centering}
\includegraphics[clip,width=1\columnwidth]{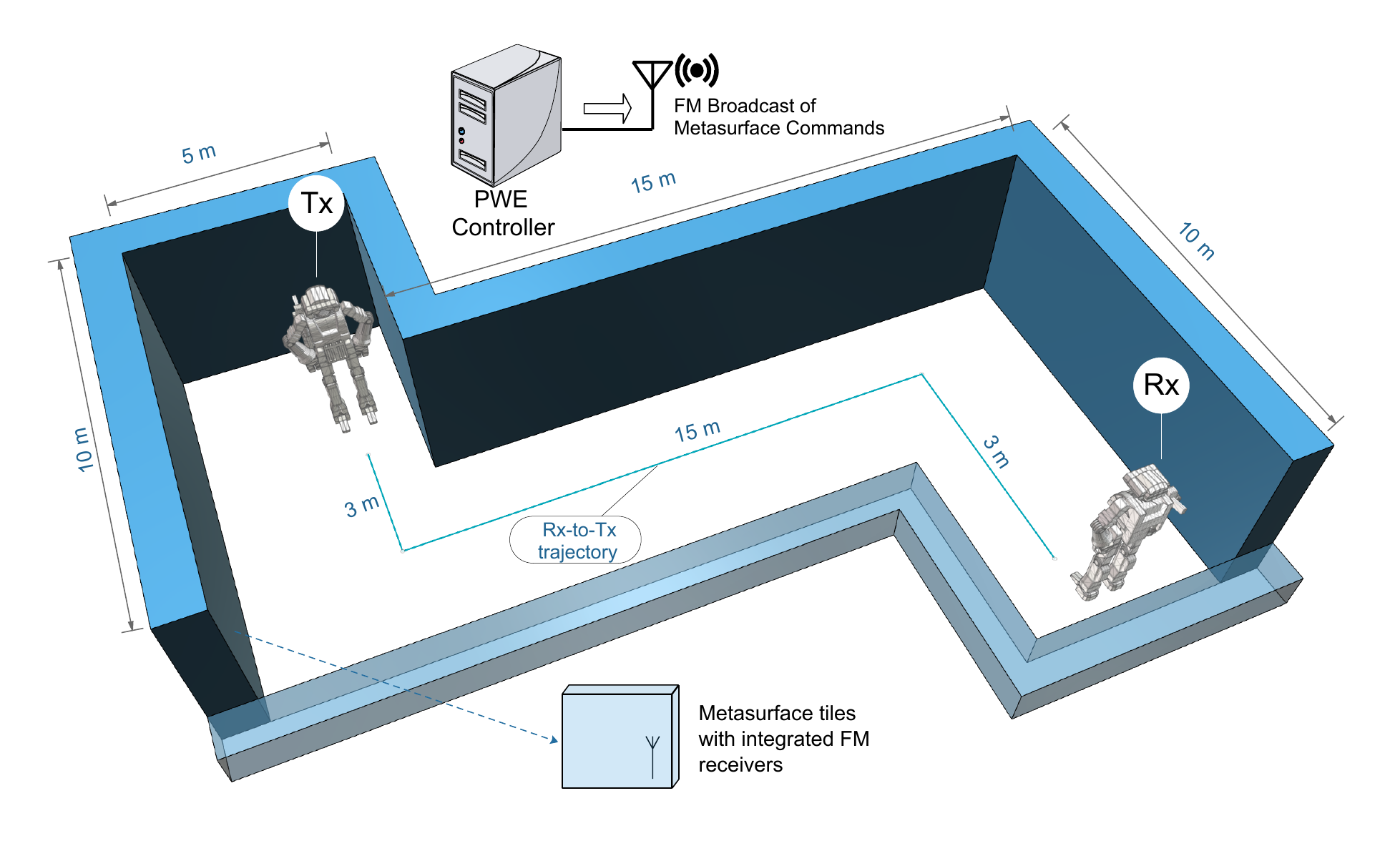}
\par\end{centering}
\caption{\foreignlanguage{english}{Overview of the employed simulation scenario.}}\label{fig:PropDoppler}
\end{figure}
We proceed to examine the setup of Fig.~\ref{fig:PropDoppler}, studying
the segment of a symmetric Z-shaped corridor (dimensions are embedded
in the Figure). All the walls and the ceiling are covered with metasurfaces
specified in Table~\ref{tab:TSimParams} and scaled for operation
in the employed mmWave frequency. Metasurfaces can be active and configured
for an EM Function, in which case they operate per their physical
design~\cite{ref78}, or be deactivated and act as
perfect reflectors. 

A PWE server sends commands to the metasurface units using an FM broadcast
channel supporting a data rate of $360$~Kbps. (We note that this
choice is made to accentuate the potential for cheap and massive control
channel connectivity for metasurfaces in the context of this tutorial).
Thus, each metasurface is equipped with a matching FM receiver. The
PWE controller serializes and broadcasts the EM Function commands
for each metasurface unit, each in the form of ``preamble, metasurface
id, EM Function id, Parameter 1, Parameter 2, ...''~\cite{ref20},
thus creating a broadcast schedule. The schedule is concluded with
a special message that notifies each metasurface to deploy its most
recently read EM Function from the FM stream. The total size of the
schedule is considered to be $360$~Kb, accounting for up to $1000$~metasurfaces
and $360$~bits per single metasurface command. Notice that each
metasurface is consider to self-host the information required to map
the broadcast EM Function identifier and the associated parameters
to the corresponding embedded circuit element states.

Two robots communicate at the mmWave range, while the Rx moves along
a well--defined trajectory with a steady speed of $1$~m/s. The
PWE is tasked with mitigating the Doppler spread, while also maximizing
the received power at the Rx. The exact nature of the robot communication
in terms of application scenario goes beyond the scope of this tutorial
and is left undefined. Notably, the Rx robot trajectory is highly
predictable. Once the robot announces its movement characteristics
(trajectory, speed) and initiation to the PWE controller, and considering
its initial park position to be known as well, the PWE controller
can infer its position at any given time moment required. Thus, the
broadcast schedule contains commands that will be matching the bot's
position at the moment when the deploy signal is emitted. Finally,
both bots have their antennas pointing upwards, in order to: i) ensure
that there is contact with metasurface units, and ii) there is no
strong LOS component, which would be unmanageable by the PWE. 
\begin{figure}
\begin{centering}
\includegraphics[width=1\columnwidth]{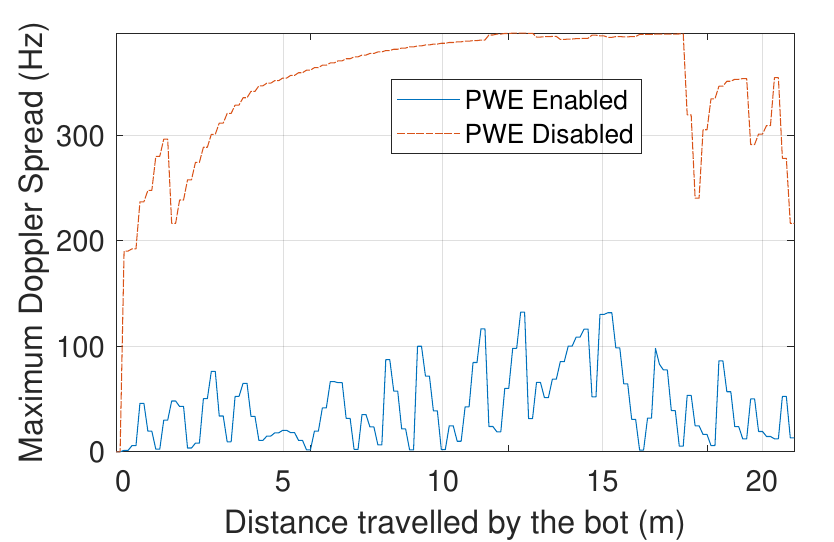}
\par\end{centering}
\caption{Maximum Doppler spread for the RX bot during its movement with PWE
enabled/disabled\foreignlanguage{english}{.}}\label{fig:DopplerResults}
\end{figure}

We proceed to examine the attained maximum Doppler spread in Fig.~\ref{fig:DopplerResults}. 

With the PWE disabled, the spread is nearly constantly at $400$~Hz.
This is especially notable in the $5$~m$\to$$10$~m traveled distance,
where the bot enters and exists the corridor respectively, and the
RF waves reach its trajectory head on.

Notably, with the PWE enabled, the spread values remain below $120$~Hz
throughout the bot's movement, and on average at $70$~Hz. A spiking
behavior is noted, with jumps from $\approx0$~Hz to $120$~Hz.
These repeat approximately every $1$~m, and essentially follow the
``freshness'' of the deployed EM Functions, which are refreshed
every $1$~sec, considering the $360$~Kb schedule size, the $360$~Kbps
FM broadcast rate and the $1$~m/s bot speed. Additional parameters
affecting the overall form of the plot are the efficiency degree of
the PWE optimization process, which can also be subject to the geometry
of the space and the placement of the tiles. 

\begin{figure}
\begin{centering}
\includegraphics[width=1\columnwidth]{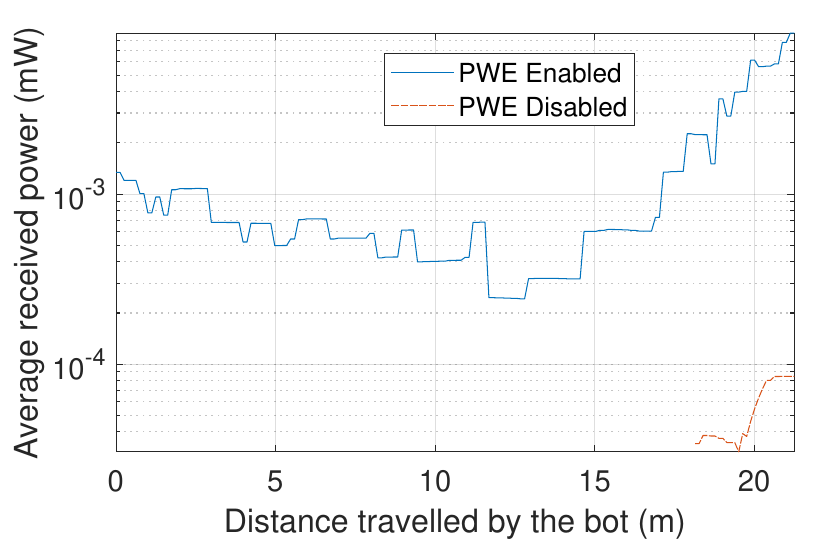}
\par\end{centering}
\caption{Average received power for the RX bot during its movement with PWE
enabled/disabled.}\label{fig:Received-power.}
\end{figure}
Figure~\ref{fig:Received-power.} shows the received power across
the bot's trajectory, with the PWE enabled and disabled. In the latter
case, the Rx bot does not receive notable power during its movement,
apart from the last part of its trajectory, when the distance to the
Tx has become minimal. This is aligned with the challenging nature
of the mmWave communications. 

When the PWE becomes enabled, the Rx bot receives power throughout
its movement. Notably, the received power drops up to a point near
the middle of corridor, and then increases once more, due to the Tx-Rx
distance reduction. This behavior at the middle of the movement is
due to the fact that less metasurface units are in close proximity
to the Rx bot at this point. As such, the EM waves cannot be as effectively
handled as, e.g., at the initial bot position, where it is effectively
surrounded by metasurface units. 

This exemplary scenario aimed at demonstrating the network/PWE interplay
and the effect on the communication quality, while also studying traits
that can only be offered by metasurface-driven setups, i.e., Doppler
spread mitigation and coverage extension in the mmWave. Notably, real-world
industrial setups may be able to offer lower-latency control than
the simple FM broadcast approach employed here, at a perhaps increased
overall cost. However, apart from the aforementioned interplay study,
it is noted that PWEs should be easily deployable and extensible at
scale, favoring their adoption and deployment in existing spaces. 

\section{Discussion and Network Research Directions}\label{sec:Discussion-and-Network}

We proceed to outline research challenges from the aspects of graph
analysis, networking, integration and machine learning. 

\subsection{Graph theoretic challenges}\label{subsec:Graph-theoretic-challenges}

As discussed in Section~\ref{subsec:A-VNE-Embedding}, the optimal
configuration of a PWE necessitates the graph exploration and exploitation
of its properties. In this context we outline two major research challenges
as follows.
\begin{figure}
\begin{centering}
\includegraphics[width=1\columnwidth]{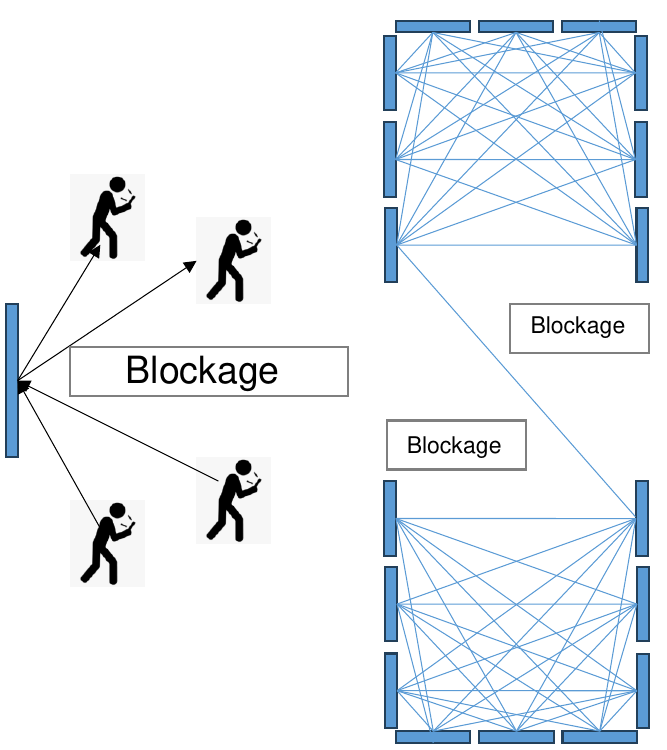}
\par\end{centering}
\caption{Visualization of the properties of symmetry (left), clique and min-cut
(right) of the PWE graph.}\label{fig:graphChallenge}
\end{figure}

\textbf{Exploitation of key-graph properties}. As shown in Fig.~\ref{fig:graphChallenge},
there exist some graph properties that call for exploitation and facilitate
the PWE configuration. In particular:
\begin{itemize}
\item The symmetry property, i.e., the existence of nodes that act as natural
or metasurface ``mirrors'', i.e., without or with the use of an
EM Function, respectively, for a set of neighboring nodes. Such nodes
allow for the deployment of a single EM Function that perfectly serves
many connectivity paths, without suffering from any loss in efficiency
that would stem from merging multiple steering EM Functions. Such
tiles can be required to act as waypoints, maximizing the number of
free tiles within the PWE after a full configuration.
\item The existence of cliques (either in the strict or in the approximate
sense) in a PWE graph is expected to be a frequently recurring property.
As tiles are placed on walls and ceilings in rectangular rooms, each
tile can be deemed connectable to all other tiles, apart from the
co-planar ones. Thus, any tile can act as a room/clique ``representative''
in an initial routing phase, and then favoring the stochastic path
exploration by delegating connectivity to any other tile around it.
\item The existence of min-cuts is also a useful property, as they define
PWE graph links (and tiles at their end-points) that provide critical
connectivity to different parts of the graph. As such, they should
undergo significant EM Function optimization, and checks for symmetry
properties, in order to server as many disparate user objectives as
possible, with minimal efficiency loss due to EM Function merging. 
\end{itemize}
Beyond these indicative properties, a thorough examination of additional
ones and their relation to the PWE configuration is considered a major
open challenge. 

\textbf{Graph path finding for PWE graphs}. Computer networking, along
with multitude other disciplines, has made extended use of path finding
algorithms. Tools such as A{*}, Dijkstra's algorithms and K-Paths
are being proven immensely useful in routing and traffic engineering.
However, these tools and many other covering cases of dynamic graphs,
have one common property: the graph nodes are re-entrant in the sense
that any path finding process that finds itself at a node will see
the same link outgoing links and weights. 

The PWE graph is peculiar in the sense that this property no longer
holds. For instance, consider a node configured for steering from
one adjacent link to another. If the path exploration process enters
the same node from an unintentional arrival link, it may find a completely
different outgoing weight (due to lower efficiency towards unintended
EM Function inputs) and even complete lack of subsequent connectivity.
As such, the existing path finding tools acquire an indicative value
in terms of their outputs. In other words, a multitude of paths are
first found and then evaluated in terms of service to the PWE configuration. 

In this aspect, a challenge is to revisit these classic aforementioned
tools and imbue them with compatibility for non re-entrant graphs.
Alternatively, new link weight definitions can be engineered in order
to increase the usefulness of the outputs of classic path finding
tools.

\subsection{Networked PWE Challenges}

As demonstrated in Section~\ref{sec:Evaluation:-A-Factory}, the
interplay between the network control and the PWE is decisive to the
performance levels that the users receive. As such, we define the
following challenges related to proactive traffic engineering, low-latency
control and network architecture. 
\begin{figure}
\begin{centering}
\includegraphics[width=1\columnwidth]{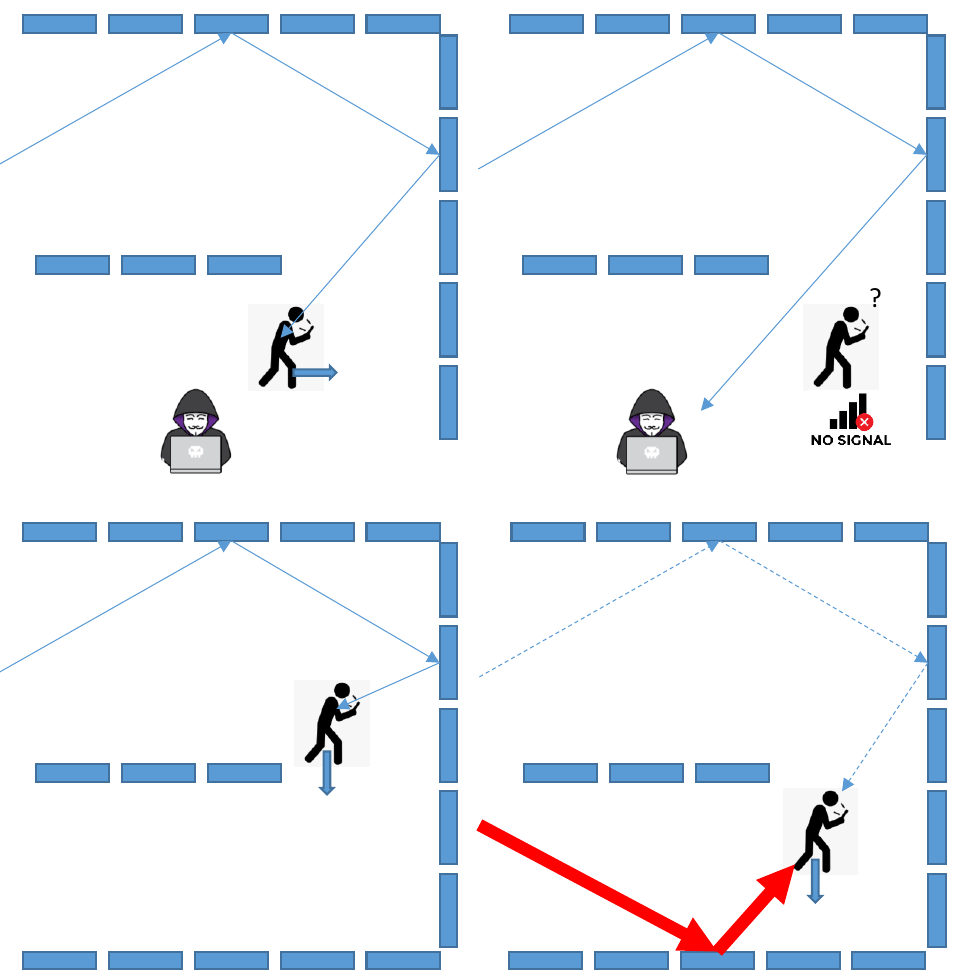}
\par\end{centering}
\caption{Visualization of PWE configuration challenges to be mitigated by
proactive traffic engineering. Top: passive eavesdropping by aligning
to the user's direction of service. Bottom: massive PWE reconfiguration
owed to user displacement.}\label{fig:graphChallenge-1}
\end{figure}

\textbf{Traffic engineering and PWEs}. Traffic engineering is the
process of distributing the traffic over a network in a manner that
fulfills performance criteria and makes the network more robust to
strain, such as denial of service attacks and outages due to hardware
impairments. In the context of PWEs, similar techniques must be enforced
to strengthen the isolation of user groups and ensure a high-level
of performance under user mobility. 

Two exemplary challenges are visualized in Fig.~\ref{fig:graphChallenge-1}:

The top part illustrates a user receiving strong signal from a tile,
while an eavesdropper has aligned himself to the user's direction
of service. A sudden displacement of the user, coupled with low network
reaction times, can expose his signal to the eavesdropper, while also
sharply reducing the user's wireless reception strength. This can
be considered a case that calls for proactive mitigation at the last-hops
of the air paths assigned to the user. First, signals must be routed
to the immediate vicinity of the user, to ensure connectivity. Second,
the paths must either follow improbable directions of arrival (e.g.,
from the ceiling), or take into account the relative position of other
users in the general vicinity. 

The bottom part illustrates a case that calls for strategic, full-path
engineering. A user that is being serviced by a major PWE path suddenly
moves to another room, where the same path attenuates beyond use.
Should the PWE routing tree adapt reactively, a considerable amount
of time may elapse before the connectivity is restored. A proactive
policy could, e.g., has already deployed secondary paths that offer
a smooth handover. 

\textbf{Interfacing with existing low-latency networking efforts}.
Managing the control and orchestration of metasurfaces within heterogeneous
network environments poses interesting technical and theoretical challenges
(HetNet~\cite{ref88}). The seamless integration
of metasurfaces requires advanced control-plane algorithms that can
efficiently coordinate between metasurface elements and existing base
stations while ensuring optimal performance within a time-sensitive
networking (TSN) framework~\cite{ref89}. This also necessitates
the development of sophisticated management protocols that can operate
in real-time. 

Another framework of interest is the Age of Information (AoI) which
promotes the concept of information freshness~\cite{ref90}.
According to it, the reading from a sensor loses its significance
following a utility function defined over the time elapsed since the
last measurement. Returning to the evaluation example of Fig.~\ref{fig:DopplerResults},
it is evident that information freshness plays a significant role
to the PWE performance, hinting prospects for an analysis that quantifies
it and an ensuing optimization. 

\textbf{PWE control architecture}. In the evaluation scenario, we
employed a cost-effective broadcast push type of PWE control employing
FM radio. Within this context, we employed a flat broadcast schedule
(also known as round-robin) to post directives to each tile. Notable,
there exist techniques for performing push schedule optimization by
increasing the number of commands for specific tiles within a single
schedule, in order to adapt to, e.g., a rapidly changing RF environment
within a specific room~\cite{ref91}.
This can decrease the control latency of push-based solutions while
retaining their cost-effectiveness. Apart from push-based control,
on-demand non-periodic broadcast scheduling techniques can be employed
as well, with the trade-off of requiring a two-way lightweight communication
with the tiles~\cite{ref92}. Additionally, the broadcast
nature of control can also be attained over wired bus architectures.

Finally, apart from a centralized control solution, there exist emerging
metasurface technologies that also enable them to act autonomically
and based on local input~\cite{ref93}, thus allowing
for a distributed (i.e., server-less) PWE control architecture. Such
metasurfaces employ autonomous impinging wave sensing~, while their
overall orchestration can be seen as a distributed control challenge
based, e.g., on consensus schemes for deducing the optimal PWE-wide
RF behavior~\cite{ref94}. 

\subsection{Standardization and Integration Challenges}

\begin{table*}[!t]
\centering{}%
\begin{tabular}{|l|l|l|}
\hline 
\textbf{Aspect} & \textbf{ETSI ISG RIS} & \textbf{3GPP (up to release 19)}\tabularnewline
\hline 
Standardization status & Dedicated ISG fully focused on RIS & No dedicated Work Item\tabularnewline
\hline 
Metasurface definition & Explicitly defined entity & Not formally defined\tabularnewline
\hline 
Architectural role & Metasurfaces as network resource & Metasurfaces as part of the RF environment\tabularnewline
\hline 
Control \& management & Control, orchestration, and lifecycle studied & No control-plane specification\tabularnewline
\hline 
Channel models & RIS-specific near-/far-field models & Generic channel models\tabularnewline
\hline 
Use cases & Communications, sensing, localization, ISAC & Discussed only at vision/study level\tabularnewline
\hline 
Deployment focus & 5G-Advanced and 6G & Mainly beyond-5G / 6G vision\tabularnewline
\hline 
Normative intent & Preparing foundation for future specs & Informative references only\tabularnewline
\hline 
\end{tabular}\caption{Comparison of ETSI ISG RIS and 3GPP Standards}\label{tab:comparison}
\end{table*}
The integration of metasurfaces into existing networks presents many
open challenges, as is evident by the state of the standards, summarized
in Table~\ref{tab:comparison}. Notice that the standardization and
regulatory framework surrounding metasurfaces is in the state of preparing
the necessary foundations (ETSI), or plainly informative (3GPP). Since
the technology is still emerging, there is significant innovation
potential towards their deployment and interoperability with existing
network infrastructures. Establishing efficient solutions and frameworks
per aspect (cf. Table~\ref{tab:comparison}) is essential for facilitating
a smooth integration process.A unified treatment from the standardization
bodies is also in need as, for instance, ETSI views metasurfaces as
a network resource, while 3GPP as a part of the RF environment only,
as shown in Table~\ref{tab:comparison}.

Nonetheless, the deployment focus on 5G-Advanced and beyond-6G networks
introduces challenges in terms of resource management and optimization
strategies. How to effectively incorporate metasurfaces into evolving
network architectures while maintaining compatibility with past and
future technologies is a critical challenge that researchers and stakeholders
must jointly address. In this aspect, it would be beneficial to align
PWEs with existing resource allocation approaches, providing extensions
of related frameworks, rather than requiring completely novel solutions. 

In this aspect we highlight the close relation between the PWE optimization
and the concept of Virtual Network Embedding (VNE)~\cite{ref95},
which also relates to the NFV standardization actions covered in Section~\ref{subsec:Katsalis},
as well as the concept of microservices~\cite{ref96}. In
general, VNE is a process within network virtualization that allows
multiple isolated virtual networks to be deployed on a single physical
infrastructure, enhancing resource efficiency. This method involves
mapping virtual components, such as virtual machines or containers,
onto physical network elements like servers and switches.

The VNE process typically begins with designing the virtual network
topology based on specific requirements. For instance, a company deploying
a cloud-based application might create a virtual network topology
consisting of three VMs connected by a virtual switch. These VMs are
then assigned to different physical servers using a virtual switch
implemented on the physical switches. The algorithms that perform
this assignment first decide whether an embedding request is admissible
to the physical infrastructure, i.e., whether there exist resources
to serve it. A greedy, coarse mapping phase is initiated, where the
most demanding virtual resources are mapped to the most under-utilized
but collocated physical resources. Stochastic search is performed
to deduce the final mapping that fulfills any performance conditions.

As noted in Sections~\ref{subsec:A-VNE-Embedding} and~\ref{subsec:Approximations-based-on},
the PWE optimization bears strong similarity to this workflow. The
user connectivity requirements (e.g., connect to a set of users and
an access point) can be viewed as an embedding request for a corresponding
virtual network, with the notable difference that the user locations
are known, acting as anchors to the VNE mapping phase. An open challenge
is to provide concrete VNE algorithmic extensions that bridge the
two concepts completely. In this aspect, separate directions can focus
on: resource optimization, which is a key benefit of VNE, as it enables
efficient use of physical infrastructure resources; scalability, allowing
new services or applications to be added without investing in additional
hardware; flexibility permitting diverse network designs on the same
infrastructure; isolation, which is crucial for security, ensuring
that a compromise in one virtual network does not affect others; simplified
network manageability by providing a logical view of the network separate
from the physical infrastructure, making it easier to monitor and
manage.

Moreover, a related concept is that of dependent microservices. This
follows the VNE rationale, yet it considers not VMs but interdependent
services that need to be: i) mapped to a physical infrastructure,
and ii) form a workflow over a data stream that traverses them. For
instance, one microservice can act as firewall, passing on the stream
to a web server. In this aspect, let EM Functions be \textquotedbl microservices\textquotedbl .
The alignment concept can be to: i) offer what the user needs (e.g.,
alter the phase, correct the polarization, etc., each can be a single
\textquotedbl microservice\textquotedbl{} that can run anywhere over
a path), and ii) keep the overall number of microservices-{}-e.g.,
steerings-{}-over each path minimal. Additionally, the EM microservices
can affect each other unintentionally (a wave \textquotedbl leaking\textquotedbl{}
can cause unintended effects), so the chosen tiles to host these \textquotedbl microservices\textquotedbl{}
(per user pair) should also have a degree of distance from microservices
that will be used for other user pairs.

Additionally, there are several challenges that need to be faced when
trying to integrate PWEs into existing radio access network (RAN)
architectures~\cite{ref97}. These challenges largely dictate
which integration scenario for PWEs is feasible. These reach from
transparent, i.e. no other network component is aware of the existence
of the PWE, to full coordination where the PWE is fully integrated.
Depending on the level of coordination different scenarios can be
supported. The main challenges involve user device tracking, switching
between a PWE and a direct connection, multiplexing multiple user
devices that are connected to the same PWE, as well as beam leaps,
where a device moves out of the LOS of the base station and connects
to PWE, leading to a sudden "jump" in its location from the perspective
of the base station. While static single device use-cases face none
of these challenges, scenarios in which multiple mobile devices with
intermittent LOS exist would face all of them. 

Accordingly, the more complex the scenario, the more tightly integrated
the PWE has to be. The main trade-off here comes in the form of protocol
extensions: While a transparent PWE can just be deployed without the
need to change the architecture at all, a fully integrated one would
require a thorough rework of the existing standards. For example,
an extension to the Open RAN alliance (ORAN) architecture would involve
the addition of at least one new interface and changes to several
of the controllers. While this is not entirely unfeasible, it is a
significant hurdle to large scale PWE deployments. As such, in any
upcoming standardization attempts, a balance needs to be struck between
which use cases to support and how many changes and extensions to
make to the specifications. 

\subsection{AI-driven Applications and Challenges}

While AI can act as a facilitator for the aforementioned challenges,
it can also act as an enabler for further applications on its own.
While potential metasurface applications such as communications, localization,
and sensing are recognized, further research is needed to explore
their practical implications and to define clear use cases that justify
the investment in the metasurface technology. Thus, novel applications
merge AI and PWEs to create systems that perform 3D monitoring of
objects--a type of RF tomography--including their material composition
in real-time~\cite{ref98}. 
\begin{figure*}[!t]
\begin{centering}
\includegraphics[width=1\textwidth]{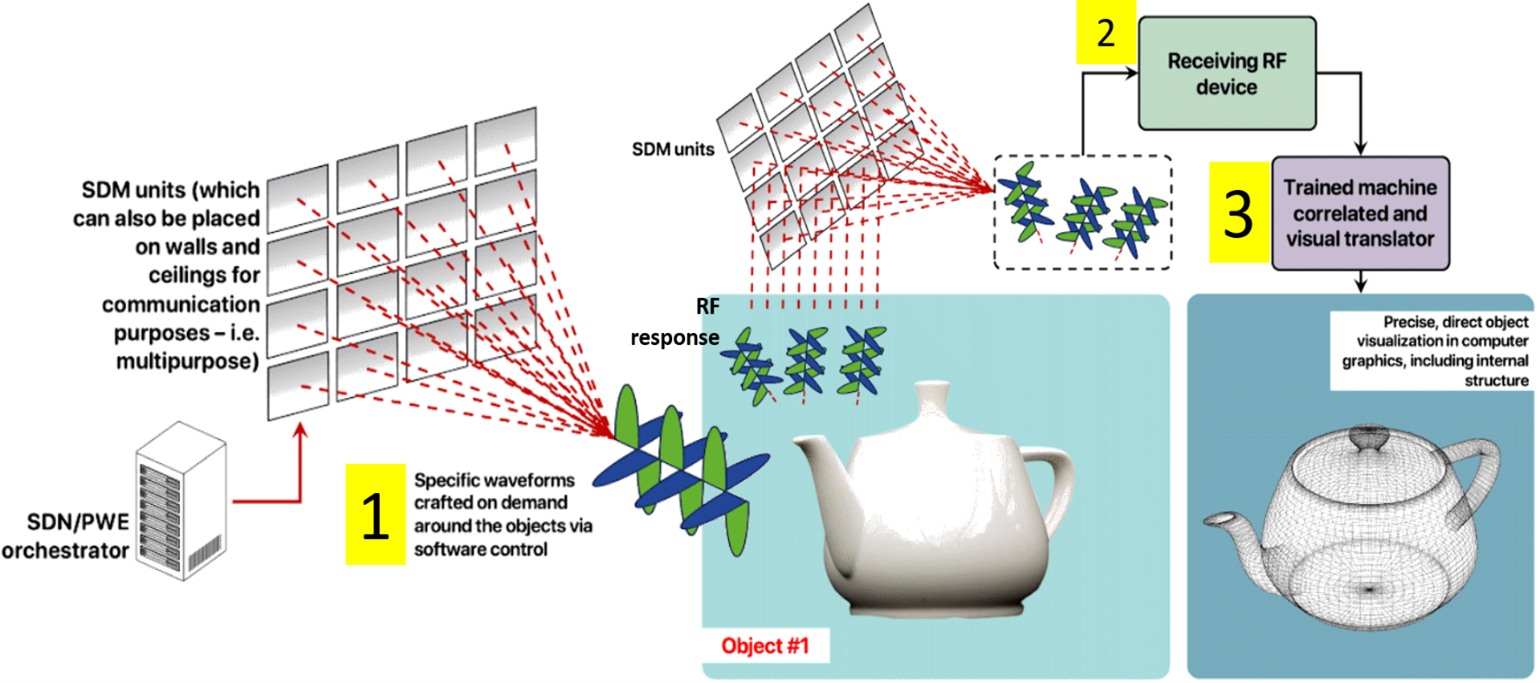}
\par\end{centering}
\caption{Principles of the envisioned AI-PWE driven RF tomography.}\label{fig:XRrf}
\end{figure*}

The operating principle is shown in Fig.~\ref{fig:XRrf}: i) Programmable
metasurfaces are employed to create an optimal radio-frequency interaction
between electromagnetic waves and an arbitrary object. ii) Radiofrequency
imaging theory principles are employed to deduce the contribution
of the wireless wavefront features (received by an array of antennas)
to the 3D visualization of an object. iii) Following a training-validation
phase, an AI system directly translates the received wavefront to
3D graphics. 

The offered advantages of this approach are that: i) Real object rotations
can be automatically transformed to rotations of its 3D representation
in real-time, without the need for computer computations, motion sensors
and lidars, and ii) The internal structure of the object is also revealed,
allowing for a type of RF tomography that is privacy preserving, has
real-time operation prospects, and does not employ ionizing radiation.
Notice that the use of AI as the RF-to-graphics converter is classified
as a non-linear process, whose fidelity bounds are not explored, contrary
to linear processes and their well-known limitations~\cite{ref99}.

\section{Conclusion}

This tutorial explored the concept of controlling metasurfaces from
a network-centric perspective. By treating metasurfaces as specialized
network components, it was shown how network theory can provide new
insights into their design, optimization, and integration with existing
infrastructure.

The paper reviewed the physical principles of metasurfaces and their
various applications, followed by an exploration of manufacturing
approaches. Subsequently, it was shown that modeling metasurfaces
as wave routers allows for describing systems of metasurfaces using
graph theory. This approach enabled the development of a performance
objective framework for optimizing these systems, with simplifications
leveraging heuristic and path-finding algorithms.

The paper also examined the integration of metasurfaces into communication
systems, discussing their overall workflow and examining their interface
with existing wireless systems from the standardization point of view,
as well as by discussing simulation prospects via existing network
discrete event simulators. An evaluation example showcasing the mitigation
of the Doppler spread in a mmWave setting provided a demonstration
of the PWE potential as well as a basic simulation setting for further
experimentation.

Finally, the paper explored future directions for research in this
field, identifying graph-theoretic challenges, networked PWE challenges,
and standardization and integration challenges. We also considered
the potential of AI-driven applications, towards application beyond
the strict context of data exchange.

\bibliographystyle{IEEEtran}
\bibliography{references}
\end{document}